\documentstyle[prb,aps, multicol,epsf]{revtex}

\begin{document}
\draft

\title{Self-consistent theory of the thermal softening and instability of
  simple crystals }

\author{  M.\ J.\ W.\ Dodgson,$^{a, b\,}$ 
V.\ B.\  Geshkenbein,$^{a, c\,}$ M.\ V.\ Feigel'man,$^{c\,}$
and G.\ Blatter$^{a\,}$}

\address{
$^{a\,}$
Theoretische Physik, ETH-H\"onggerberg, CH-8093  Z\"urich, Switzerland}

\address{
$^{b\,}$
Theory of Condensed Matter Group,
Cavendish Laboratory, Cambridge, CB3 0HE, UK}

\address{$^{c\,}$L. D. Landau Institute for Theoretical Physics,
  117940 Moscow, Russia}

\date{July 5, 2000}
\maketitle
\begin{abstract}

We consider the thermal softening of crystals due to anharmonicity. 
Self-consistent methods find a maximum temperature for a stable
crystal, which gives an upper bound to the melting temperature. 
Previous workers have shown
that the self-consistent harmonic approximation (SCHA) gives
misleading results for the thermal stability of crystals. The reason is that
the most important diagrams in the perturbation expansion around harmonic 
theory are not included in the SCHA.
An alternative approach is to solve a self-consistent Dyson equation (SCA)
for a selection of diagrams, the simplest being a (3+4)-SCA. However, 
this gives an unsatisfactory comparison to numerical results on the
thermal and quantum melting of two-dimensional (2D)
Coulomb-interacting particles (equivalent to vortex-lattice melting in two and 
three dimensions). 
We derive an
improved self-consistent method, the two-vertex-SCHA, which gives much better
agreement to the simulations.
Our method allows for accurate
calculation of the thermal softening of the shear modulus for 2D crystals and
for the lattice of vortex-lines
in type-II superconductors.

\end{abstract}


\begin{multicols}{2}
\narrowtext

\section{Introduction}\label{sec:intro}

\begin{figure}
\centerline{\epsfxsize= 6.5cm\epsfbox{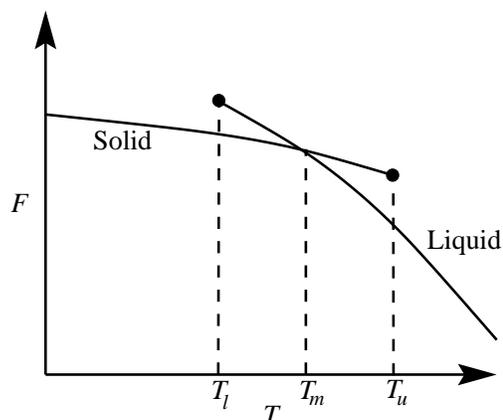}}
\caption{ Schematic diagram of the free energy as a function of temperature
  for a first-order phase transition. The thermodynamic phase is the one of
  lowest free energy, and the possibility of a (meta-)stable solution of a 
  phase  exists beyond the temperature of the transition. \label{fig:1}}
\end{figure}

It was realized long ago on theoretical grounds that most melting transitions
should be first order,\cite{Landau} although exceptions include the
possibility of defect-mediated melting in two-dimensional (2D) 
crystals\cite{HN}
and
the mean-field transition between the superconducting vortex lattice and the
normal state in type-II superconductors.\cite{Abrikosov} 
As with most first-order transitions, there is no good simple theory
of melting to compare with our theoretical understanding of many continuous
phase transitions: A first-order transition occurs when the free energy of one
phase as a function of temperature crosses the free energy of another phase 
(see Fig.~\ref{fig:1}). 
These two phases are qualitatively different and generally demand
theoretical treatments with distinct approximations, such that a detailed
comparison of free energies is not possible. The solid-liquid transition is a
prime example of this problem. We can treat the solid phase well within
elasticity theory, but this is not a useful description of the liquid.

A common approach to this problem is to concentrate on the solid phase and
calculate its stability limit\cite{Pietronero} (a complementary method is to 
treat the liquid, e.g. within density functional theory,\cite{Ramakrishnan} 
and find the lowest temperature for the liquid phase to exist).
A plausible mechanism for the crystal instability comes from anharmonic 
effects that may soften
the lattice when thermal fluctuations are large. To this end one can develop a
self-consistent harmonic approximation (SCHA),\cite{Pietronero,Kleinert} 
where an effective harmonic
theory is used that better approximates the true anharmonic system for finite
thermal fluctuations. We can then identify the limit to a
self-consistent solution with the instability temperature of the solid.
While it is unclear whether
or not the instability point $T_u$ is close to the melting temperature $T_m$,
it does give an upper bound (see Section~\ref{sec:SCHA}), 
and represents the first step to understanding
the possible mechanisms of melting for a given system.

An example of this hunt for the anharmonic instability point occurred several
years ago in the context of 2D melting. An early SCHA calculation by Platzman 
and Fukuyama\cite{PlatzmanFukuyama} 
for a 2D electron system found a dramatic stiffening of the
Wigner crystal as a function of temperature, and the anharmonic instability
point was reached well above the dislocation-mediated melting
temperature. Later, Fisher considered anharmonic corrections from a 
perturbation expansion around the harmonic theory, and could show that the
lowest-order temperature correction to the shear modulus was
downwards.\cite{Fisher2} 
The opposite result of the SCHA calculation comes from the
fact that it misses contributions from odd anharmonicities. A series of papers
by Lozovik and coworkers\cite{Lozovik} 
emphasized this fact, and developed an alternative
self-consistent procedure from the Dyson equation for the two
lowest order diagrams of the expansion. This alternative method was
named the (3+4)-self consistent
approximation (SCA). (The SCA is distinguished from the SCHA in that
self-consistency comes from the Dyson resummation of a set of skeletal
diagrams in the perturbation theory, rather than by using an effective harmonic
theory to take averages over the true Hamiltonian.)
An important result of
Lozovik's work was that for long-range interactions in 2D, the anharmonic
instability point lies below the dislocation mediated melting
temperature, implying a first-order rather than continuous transition.

Our own interest in the melting problem originates in studies of the vortex
system in high-$T_c$ superconductors (see Ref.~\onlinecite{review} for a 
review.) There is excellent experimental evidence for a first-order
melting transition of the vortex lattice in clean crystals of
Bi$_2$Sr$_2$CaCu$_2$O$_8$ (BSCCO)\cite{Zeldov} and
YBa$_2$Cu$_3$O$_{7-\delta}$ (YBCO),\cite{Welp,Schilling}
with a latent heat and magnetization step consistent with our
theoretical models of the vortex system.\cite{Dodgson1}
However, the detailed understanding of where and why this transition 
takes place has so far relied on numerical
simulations.\cite{Nordborg,XHu,Koshelev,Nguyen,SasikStroud,HuMac} 
With this in mind, we have tried to apply the above
analytic extensions to elasticity theory within the London
model\cite{Brandt_Elastic} 
to derive the instability point of the vortex crystal. By comparing our
results to the numerical simulations of Ref.~\onlinecite{Nordborg} we have
found, in agreement with Lozovik, that the SCHA seriously underestimates the 
thermal softening of the vortex lattice (see Section~\ref{sec:VL}). However, 
application of the
diagrammatic (3+4)-SCA did not lead to a good quantitative comparison, and 
so we have developed 
an improved self-consistent method that includes all diagrams in the (3+4)-SCA 
and the SCHA. Because the improved method uses an effective harmonic theory,
but is equivalent to the Dyson equation for all two-vertex
diagrams, we call it the two-vertex-SCHA. 
The successful comparison of this new method to numerical
results for the vortex lattice forms the main result of this paper.

The article is structured as follows. We first define the harmonic
approximation for a generalized crystal. Then in Section~\ref{sec:SCHA}
we describe the
self-consistent harmonic approximation and its justification within a
variational approach. In Section~\ref{sec:pert}
we review the perturbation expansion about
the harmonic approximation, look at the diagrams corresponding to the SCHA and
see why this approximation can give such misleading results. The alternative
(3+4)-SCA of Lozovik {\it et al.} 
is also introduced, and we discuss the advantages and
disadvantages over the SCHA. After learning the diagrammatic details of these
self-consistent approximations, in Section~\ref{sec:new}
we  introduce our improved
self-consistent method, the two-vertex-SCHA, 
that includes all diagrams in both the SCHA, (3+4)-SCA
and more, while keeping some of the simplicity in treatment of the
SCHA. Finally in  Section~\ref{sec:real} 
we apply this improved method to some physical
problems. First we consider the thermal melting in two dimensions of the
``one-component plasma'' (with $\ln(R)$ interactions) and the Wigner crystal
(with $1/R$ interactions). In contrast to the (3+4)-SCA, our results from the
two-vertex-SCHA leave open the possibility of dislocation-mediated melting.
We also calculate 
the quantum melting of 2D bosons, equivalent to thermal melting of vortex
lines. We find a significant thermal softening of the vortex lattice, which
could be responsible for the observed peak effect\cite{PeakEffect} 
in critical current close to the melting transition in YBCO.

\section{Review of Harmonic theory}

We consider a classical $d$-dimensional crystal at finite temperature, where
the particle positions make small fluctuations $\{{\bf u}_\mu\}$
about their equilibrium sites $\{{\bf R}_\mu\}$. All thermodynamic
properties are controlled by the partition function,
\begin{equation}
  Z=\int\prod_\mu d^du_\mu\,\, e^{-H[{\bf u}_\mu]/T},
\end{equation}
where the Hamiltonian $H[{\bf u}_\mu]$ gives the energy increase from the
ground state for a given configuration of displacements.\cite{GSE}
We work in Fourier space (which diagonalizes a harmonic Hamiltonian),
and define,
\begin{equation}
 {\bf u}_{\bf k}=\frac{1}{n}
\sum_{{\bf R}_\mu} {\bf u}_\mu e^{-i{\bf k}\cdot{\bf R}_\mu},
\end{equation}
where $n$ is the particle density. The inverse transform is an integral 
over the Brillouin zone,
\begin{equation}
 {\bf u}_\mu  =\int_{\rm BZ}\frac{d^dk}{(2\pi)^d}\,\,
{\bf u}_{\bf k}e^{i{\bf k}\cdot{\bf R}_\mu}.
\end{equation}
In the harmonic approximation (HA) we 
expand the Hamiltonian to second order in displacements,
\begin{equation}\label{eq:h2}
  H[{\bf u}_\mu]=H_2[{\bf u}_\mu] + {\cal O}(u^3), 
\end{equation}
with 
\begin{eqnarray}
  H_2[{\bf u}_\mu] &=&\frac{1}{2n^2}\sum_\mu\sum_\nu 
\Phi_0^{\alpha\beta}({\bf R}_\mu-{\bf R}_\nu) u_\mu^\alpha u_\nu^\beta,\\
&=&\frac{1}{2}
\int_{\rm BZ}\frac{d^dk}{(2\pi)^d}
\Phi_0^{\alpha\beta}({\bf k})u^\alpha_{\bf k}u^\beta_{-{\bf k}}. 
\end{eqnarray}

The elastic matrix is defined by 
\begin{equation}\label{eq:elastic}
 \Phi_0^{\alpha\beta}({\bf k})= 
L^d\left.\frac{\delta^2 H}{\delta u^\alpha_{\bf k}
\delta u^\beta_{-{\bf k}}}\right|_{u=0}.
\end{equation}
Here $L^d$ 
is the total volume, which determines the discretization of $k$-space.
We define the harmonic 
Green's function as the inverse of the elastic matrix,
$G_0^{\alpha\beta}({\bf k})= (\Phi_0^{-1})^{\alpha\beta}({\bf k})$.

The HA, being a quadratic theory, is completely solvable, with the size of 
fluctuations given by the equipartition theorem. The 
harmonic propagator is the thermal average,
\begin{eqnarray}\label{eq:equiofk}
  \langle u^\alpha_{{\bf k}_1}u^\beta_{{\bf k}_2}\rangle_0
&=&Z_0^{-1} \int \Pi_{{\bf k}'} \, d^d u_{{\bf k}'}\,
u^\alpha_{{\bf k}_1}u^\beta_{{\bf k}_2}
e^{-H_2[{\bf u}_{{\bf k}'}]/T}\nonumber\\
&=&T\, \delta^d({\bf k}_1+{\bf k}_2)G_0^{\alpha\beta}({\bf k}_1),
\end{eqnarray}
where $Z_0$ is the partition function for the harmonic Hamiltonian $H_2$.
For dimensions $d>2$ a useful measure of the fluctuations is the mean-square
displacement of each particle (in 2D we need to be careful about
divergences at long wavelengths, see Appendix~\ref{ap:nn}). 
This is just the integral over ${\bf k}$ of the 
propagator,
\begin{equation}\label{eq:equi}
  \langle u^2\rangle_0 = \langle u^\alpha({\bf R})u^\alpha({\bf R})\rangle_0
= T\int_{\rm BZ} \frac{d^d k}{(2\pi)^d} G_0^{\alpha\alpha}({\bf k}). 
\end{equation}

\section{The self-consistent harmonic approximation}\label{sec:SCHA}

From Eq.~(\ref{eq:equi}) we see how the mean displacements increase with
temperature. At some stage the anharmonic corrections to Eq.~(\ref{eq:h2}) 
become important. One way to treat these anharmonic effects is to take the
quadratic form as a trial Hamiltonian, 
\begin{equation}\label{eq:trial}
  H_t[{\bf u}_{\bf k}] 
=\frac{1}{2}
\int_{\rm BZ}\frac{d^dk}{(2\pi)^d}
\Phi_t^{\alpha\beta}({\bf k})u^\alpha_{\bf k}u^\beta_{-{\bf k}},
\end{equation}
with the elastic matrix as a set of
variational parameters. We can then use the general inequality\cite{Feynman}
for the free energy $F=-T\ln{Z}$,
\begin{equation}\label{eq:ineq}
  F \le F_t +\langle H-H_t\rangle_t, 
\end{equation}
so that for a given trial Hamiltonian we get the best approximation
to the free energy by minimizing the right hand side. It is straightforward to
show that for the quadratic form
this minimization is satisfied by the effective elastic matrix,
\begin{equation}\label{eq:eff}
 \Phi_t^{\alpha\beta}({\bf k})= 
L^d\left\langle\frac{\delta^2 H}{\delta u^\alpha_{\bf k}
\delta u^\beta_{-{\bf k}}}\right\rangle_t,
\end{equation}
where the average is over the distribution defined by the trial
Hamiltonian (\ref{eq:trial}) at the given temperature. This must
be solved self-consistently, which defines the SCHA.

Let us consider the general case of a system of particles with pairwise
interactions,
\begin{equation} \label{eq:pairwise}
  H = \frac{1}{2} \sum_{\mu\ne\nu} V({\bf R}_\mu + {\bf u}_\mu 
- {\bf R}_\nu - {\bf u}_\nu),  
\end{equation}
where ${\bf R}_\mu$ are the ideal crystal positions with nearest neighbor
separation $a$. Eq.~(\ref{eq:eff}) then
becomes,
\begin{equation}\label{eq:effpair}
 \Phi_t^{\alpha\beta}({\bf k})= 
n\sum_{\mu\ne 0} (1-\cos{\bf k}\cdot{\bf R}_\mu)
\left\langle
\left.\frac{\partial^2 V}{\partial r^\alpha \partial r^\beta}
\right|_{ {\bf r}= {\bf R}_\mu + {\bf u}_\mu- {\bf u}_0}
    \right\rangle_t.
\end{equation}
This average over the second derivative of the potential is conveniently
written in terms of the displacement fluctuations by Fourier transforming the
interaction potential,
\begin{eqnarray}
  \left\langle
\left.\frac{\partial^2 V}{\partial r^\alpha \partial r^\beta}
\right|_{ {\bf r}= {\bf R}_\mu + {\bf u}_\mu- {\bf u}_0}    \right\rangle_t
&&\\
&&\hspace{-3.8cm}
=-\int\frac {d^dq}{(2\pi)^d} q^\alpha q^\beta V(q) e^{i{\bf q}\cdot{\bf R}_\mu}
e^{-\frac 1 2 q^\alpha q^\beta 
\langle (u^\alpha_\mu-u^\alpha_0)(u^\beta_\mu-u^\beta_0)\rangle}.\nonumber
\end{eqnarray}
Because the fields in a harmonic potential will have Gaussian fluctuations,
we have used the property $\langle e^{ikx}\rangle=
e^{-\frac 1 2 k^2\langle x^2\rangle}$ for a Gaussian distributed variable
$x$. Although not necessary 
for calculations, the formalism
simplifies greatly if we can make the approximation that 
$\langle (u^\alpha_\mu-u^\alpha_0)(u^\beta_\mu-u^\beta_0)\rangle =(2/d) 
\delta_{\alpha\beta} \langle u^2\rangle$, which means ignoring correlations
between the fluctuations of different particle positions. In this case we
find,
\begin{eqnarray}\label{eq:effsimple}
 \Phi_t^{\alpha\beta}({\bf k})&=& 
n^2\sum_{{\bf Q}_\mu } ({\bf Q}_\mu+{\bf k})^\alpha ({\bf Q}_\mu+{\bf
  k})^\beta f(|{\bf Q}_\mu+{\bf k}|)\nonumber\\
&&\hspace{1cm}-Q_\mu^\alpha Q_\mu^\beta f(Q_\mu),
\end{eqnarray}
where ${\bf Q}_\mu$ are the reciprocal lattice vectors, and 
$f(q)=V(q)e^{-q^2\langle u^2\rangle/d}$.
The fluctuations $\langle u^2\rangle$ are calculated from the equipartition
theorem (\ref{eq:equi}) but with the Green's function given as the inverse of
the effective elastic matrix. 
Thus within this homogeneous approximation, we have turned the problem of
finding $\Phi_t^{\alpha\beta}({\bf k})$ for all ${\bf k}$ to a self-consistent
equation for a single parameter, $\langle u^2\rangle$. 
From (\ref{eq:effsimple})
it is clear that at large 
$\langle u^2\rangle$ the elastic matrix will be suppressed from the zero
temperature limit, which in turn increases $\langle u^2\rangle$ above the
harmonic expectation. It is this feature which drives an anharmonic
instability of the crystal phase at a certain temperature.

\begin{figure}
\centerline{\epsfxsize= 8cm\epsfbox{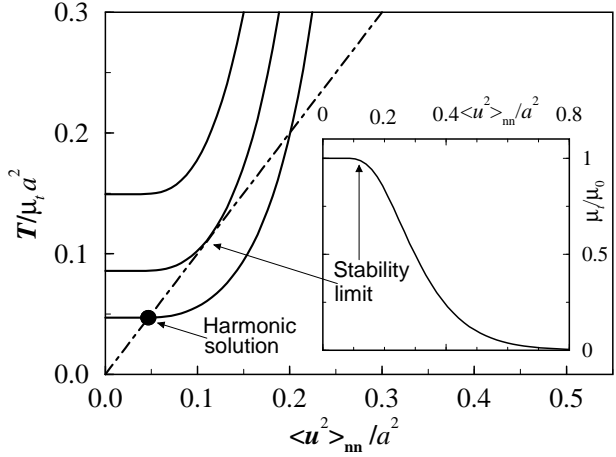}}
\caption{ 
  Illustration of the SCHA for a 2D Coulomb system (see Section~\ref{sec:2D}).
  The inset shows the shear modulus $\mu_t$ of the 2D Coulomb lattice 
  as a function of nearest-neighbor displacements
$\langle u^2\rangle_{\rm nn}$, 
  as given in Eq.~(\ref{eq:c66ofu2}).
  We show graphical solutions to the equipartition result,   
  $\langle u^2\rangle_{\rm nn}\sim T/\mu_t$ 
  for three different temperatures. At high
  temperatures there are no self-consistent solutions. Note the 
  non-perturbative form of $\mu_t$ at small fluctuations: The curve cannot be
  approximated by a power series in $\langle u^2\rangle_{\rm nn}$. 
  In fact, further
  investigation of the diagrammatics shows that this result severely
  underestimates the thermal softening of the lattice.}
\label{fig:2D_c66ofu2}
\end{figure}

The SCHA in the ``independent fluctuations'' approximation has an interesting
feature for the case of Coulomb interactions ($1/r$ in three dimensions, $\ln
r$ in two dimensions etc.). By definition the Laplacian of
a Coulomb interaction is zero, and one can show that the renormalized 
interaction, and therefore the effective elastic matrix,
is a non-perturbative function of $\langle u^2\rangle$. This is shown
explicitly for the case of logarithmic interactions in two dimensions in
Appendix~\ref{ap:log}. For the 2D case, the single particle fluctuations 
diverge (see Appendix~\ref{ap:nn}) even at low temperatures where the shear 
modulus remains non-zero.\cite{HN} However,
the shear modulus 
$\mu_t=\lim_{k_y\rightarrow 0}\left[\Phi^{xx}_t(k_y)/{k_y}^2\right]$ is 
dominated by the nearest-neighbor fluctuations
$\langle u^2\rangle_{\rm nn}=
\frac 12 \langle \left| {\bf u}({\bf R}_\mu)- 
{\bf u}({\bf R}_{\mu+1})\right|^2\rangle$; the formula for 
$\mu_t(\langle u^2\rangle_{\rm nn})$
is given in Section~\ref{sec:2D}
and shown in the inset of Fig.~\ref{fig:2D_c66ofu2}.
This allows us to illustrate
the SCHA in its simplest form. For this 2D Coulomb system we can approximate 
the equipartition result as $\langle u^2\rangle_{\rm nn}
\sim T/\mu_t$ (i.e. we ignore
compression modes, and assume no dispersion). The two
equations relating $\langle u^2\rangle_{\rm nn}$ and $\mu_t$
can then be solved graphically for any temperature
as in Fig.~\ref{fig:2D_c66ofu2}, and the maximum
value of $\langle u^2\rangle_{\rm nn}$ 
for a stable self-consistent solution is easily
found. In this case we find the large value of 
$\langle u^2\rangle_{\rm nn}/a^2\approx (0.34 )^2$ at $T_u$. 

The following argument indicates that this instability point in the SCHA is a
rigorous upper bound on the true stability limit of the crystal phase. If
there is no solution to the SCHA then by definition there is no stationary
value of the RHS of Eq.~(\ref{eq:ineq}), which implies that this RHS
is an unbounded function of trial parameters. Because of the inequality, there 
can be no finite free energy $F$, and no equilibrium phase
corresponding to the exact Hamiltonian $H[{\bf u}_{\bf k}]$. Therefore the
only possible equilibrium phase is one without a well-defined
mapping from particle positions to reference lattice points. In this case the 
Fourier transforms ${\bf u}_{\bf k}$ are not the relevant variables of the 
system, and  the liquid phase is still allowed.

Unfortunately, it is known that the SCHA is not always a reliable
approximation, and we referred in the introduction to an example 
where it drastically underestimates the thermal softening of the 
lattice.\cite{PlatzmanFukuyama} 
To understand why this is so we must consider which diagrams the SCHA 
represents in a perturbation expansion around the harmonic limit, and this is 
done in Section~\ref{sec:diagsSCHA}.

\section{Perturbation theory beyond the harmonic approximation}\label{sec:pert}

\begin{figure}
\centerline{\epsfxsize=8.5cm\epsfbox{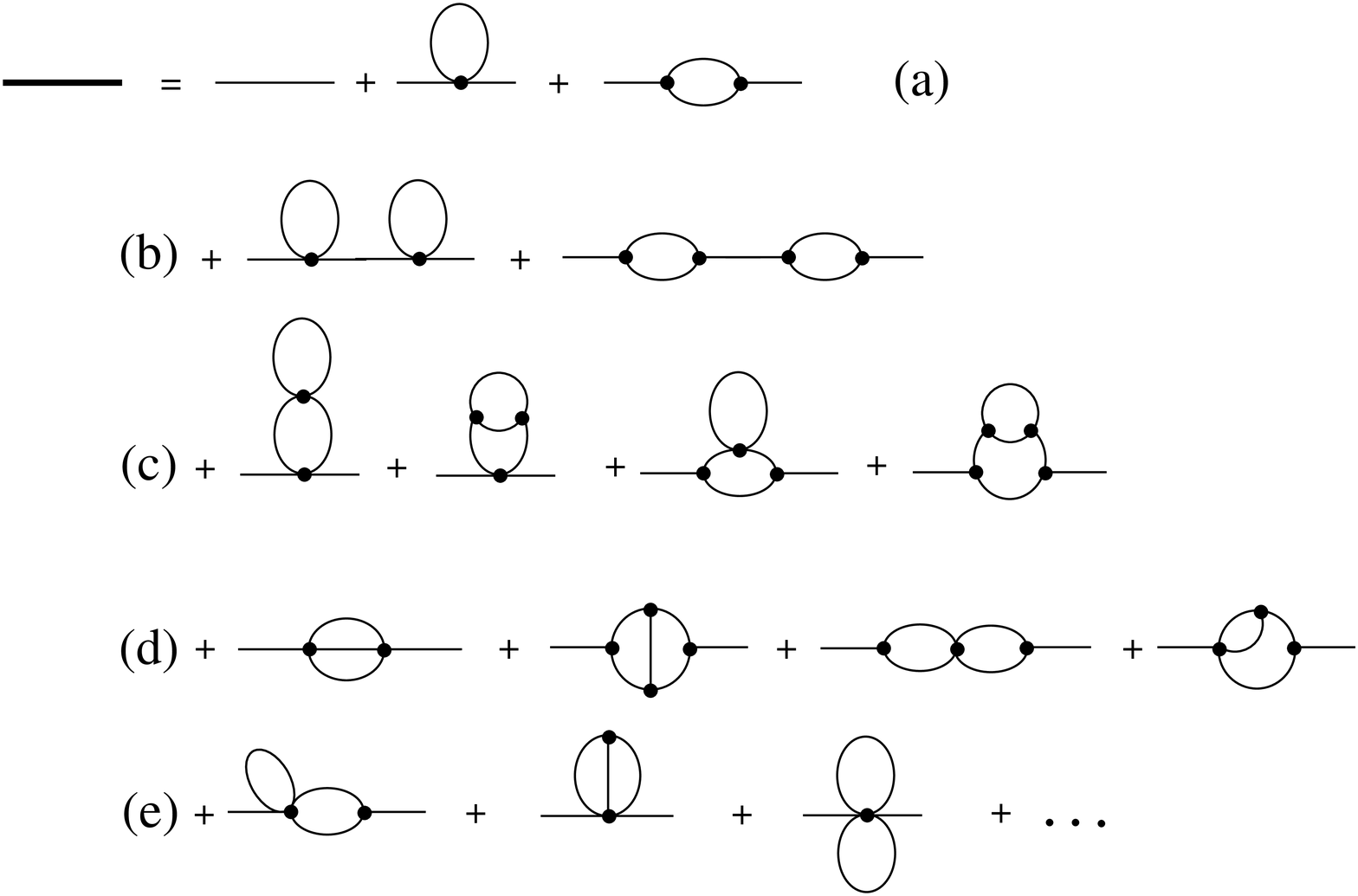}}
\caption{ The diagrammatic expansion of the propagator as described in the
  text, to order $T^3$. 
  We have drawn the diagrams in four groups. Group (a)
  represents the expansion to order $T^2$, which only involves third and
  fourth order vertices. The remaining groups are all of order $T^3$.
  Group (b) shows the reducible diagrams that are trivial to resum (see text).
  Group (c) shows the irreducible
  diagrams to this order that can still be generated by a self-consistent
  resummation of the lowest order diagrams. Group (d) shows the diagrams that
  involve only third and fourth order vertices, yet are topologically new to
  this order. Group (e) shows the first diagrams that include fifth and sixth
  order vertices. }
\label{fig:diagrams}
\end{figure}

We now consider the standard perturbation expansion about the
HA.\cite{Choquard}
 The exact
partition function is,
\begin{equation}
  Z=\int\prod_{\bf k} d^du_{\bf k}\,\, 
e^{-\left\{H_2[{\bf u}_{\bf k}]
+H'[{\bf u}_{\bf k}]\right\}/T},
\end{equation}
where $H'[{\bf u}_{\bf k}]=H[{\bf u}_{\bf k}] - H_2[{\bf u}_{\bf k}]$.
As usual we expand the exponential in $H'$,
\begin{eqnarray}
  Z&=&Z_0 \\
&&+\sum_{n=1}^\infty\frac{(-1)^n}{n!}
\int\prod_{\bf k} d^du_{\bf k}\,\, 
\left(\frac{H'[{\bf u}_{\bf k}]}{T}\right)^n
e^{- H_2[{\bf u}_{\bf k}]/T}.\nonumber
\end{eqnarray}
With the same expansion, the propagator is,
\begin{eqnarray}
  \langle u^\alpha_{{\bf k}}u^\beta_{-{\bf k}}\rangle&=&
Z^{-1}
\int\prod_{{\bf k}'} d^du_{{\bf k}'}\,\,
 u^\alpha_{{\bf k}}u^\beta_{-{\bf k}}
e^{-\left\{ H_2[{\bf u}_{{\bf k}'}]
+H'[{\bf u}_{{\bf k}'}]\right\}/T}\nonumber\\
&=&
\langle u^\alpha_{{\bf k}}u^\beta_{-{\bf k}}\rangle_0\\
&&
+\sum_{n=1}^\infty\frac{(-1)^n}{n!}
\left\langle\left\|  u^\alpha_{{\bf k}}u^\beta_{-{\bf k}}\,
\left(\frac{H'[{\bf u}_{{\bf k}'}]}{T}\right)^n\right\|\right\rangle_{c,G_0}.
\nonumber\end{eqnarray}
The brackets $\left\langle\left\|\ldots\right\|\right\rangle_{c,G_0}$ denote
a cumulant average in the HA, that is, using Wick's theorem
the $u$ fields must be paired up to give harmonic Green's functions, $G_0$,
and all terms with disconnected pairings are ignored, as these cancel exactly
with the expansion of the partition function $Z$ in the denominator. 
To establish the terms in this series at each order in $T$, we
need to separate each order in $u$ of the Hamiltonian,
\begin{equation}
 H'[{\bf u}_{\bf k}]=H_3[{\bf u}_{\bf k}] + H_4[{\bf u}_{\bf k}]+ 
H_5[{\bf u}_{\bf k}]\,+\,\ldots
\end{equation}
where $H_m[{\bf u}_{\bf k}]$ contains all terms of order $u^m$, and will lead
to a vertex in the diagrammatic expansion with $m$ legs.
Explicitly,
\begin{equation}\label{eq:Hm}
  H_m=\frac{1}{m!}\int_{BZ}\prod_{i=1}^m 
  \frac{d^dk_i}{(2\pi)^d}\,\,
  A^{\lambda_1\ldots\lambda_m}({\bf k}_1,\ldots{\bf
    k}_m)\,
u^{\lambda_1}_{{\bf k}_1}\ldots u^{\lambda_m}_{{\bf k}_m}.
\end{equation}
The crystal symmetry of the ground state lattice implies that the interaction
tensor conserves pseudomomentum,
\begin{eqnarray} \label{eq:AtoPhi}
   A^{\lambda_1\ldots\lambda_m}({\bf k}_1,\ldots{\bf k}_m)
&\equiv& (L^d)^m 
\left.\frac{\delta^m H} {\delta u^{\lambda_1}_{{\bf k}_1}\ldots 
\delta u^{\lambda_m}_{{\bf k}_m}}\right|_{u=0}
\\
&&\hspace{-1.3cm}=
 \Phi^{\lambda_1\ldots\lambda_m}({\bf k}_1,
\ldots,{\bf k}_{m-1})
\Delta^d({\bf k}_1+\ldots{\bf k}_m),\nonumber
\end{eqnarray}
where $\Delta^d({\bf k})=\sum_{{\bf Q}_\mu} \delta^d({\bf k}+{\bf Q}_\mu)$.
The important point is that in the HA the expectation of the propagator 
$\langle u_{\bf k}u_{-{\bf k}}\rangle_0$ is proportional to $T$, so that the
average value of $H_m$ is of order $T^{\frac m 2}$. 
We can write the perturbation expansion as,
\begin{eqnarray}\label{eq:pert}
  \langle u^\alpha_{{\bf k}}u^\beta_{-{\bf k}}\rangle&=&
\langle u^\alpha_{{\bf k}}u^\beta_{-{\bf k}}\rangle_0\\
&&
+\sum_{n=1}^\infty\frac{(-1)^n}{n!}
\left\langle\left\|  u^\alpha_{{\bf k}}u^\beta_{-{\bf k}}\,
\left(
\sum_{m=3}^{\infty}\frac{H_m}{T}\right)^n\right\|\right\rangle_{c,G_0},
\nonumber
\end{eqnarray}
and then take all possible pairings using Wick's theorem. We can
represent this series in diagrams and the terms to
order $T^3$ are shown in Fig.~\ref{fig:diagrams}. 
The order of each diagram (in $T$) is counted by adding up
the number of lines, and subtracting the number of vertices.
Notice that to order $T^2$, only $H_3$ and
$H_4$ contribute, while to order $T^3$ we also have diagrams including 
$H_5$ and   $H_6$. This is a general rule: the terms $H_{2m}$ and 
$H_{2m-1}$ first appear to order $T^{m}$.

We now make a resummation that allows us to express the full propagator $G$
in terms of the harmonic propagator $G_0$ and a self energy $\Sigma$.
Note that, because we do not
include disconnected diagrams, each $u_{\bf k}$ in (\ref{eq:pert}) must join
with a $u_{{\bf k}'}$ in $H'$ when using Wick's theorem. This means we can
rewrite (\ref{eq:pert}) in the form,
\begin{eqnarray}\label{eq:pert2}
  \langle u^\alpha_{{\bf k}}u^\beta_{-{\bf k}}\rangle&=&
\langle u^\alpha_{{\bf k}}u^\beta_{-{\bf k}}\rangle_0\\
&&
+
\langle u^\alpha_{{\bf k}}u^{\lambda_1}_{-{\bf k}}\rangle_0
\langle  u^{\lambda_2}_{{\bf k}}u^\beta_{-{\bf k}}\rangle_0\nonumber\\
&&\hspace{-0.5cm} \times
\sum_{n=1}^\infty\frac{(-1)^n}{n!}
\left\langle\left\|  \frac{\delta^2}{\delta u^{\lambda_1}_{-{\bf k}}
\delta u^{\lambda_2}_{{\bf k}}}
\left(
\sum_{m=3}^{\infty}\frac{H_m}{T}\right)^n\right\|\right\rangle_{c,G_0}.
\nonumber
\end{eqnarray}
Inspection of Fig.~\ref{fig:diagrams} shows that
many possible terms will be trivial ``copies'' of lower order
terms. These are the reducible diagrams, for instance group (b) in
Fig.~\ref{fig:diagrams}, which have the property that when we group all
of the reducible diagrams corresponding to a given irreducible diagram, they
form a resummable series. We can write this in the compact form
\begin{equation}\label{eq:greens}
  G^{\alpha\beta}({\bf k})=G_0^{\alpha\beta}({\bf k})+
 G_0^{\alpha\lambda_1}({\bf k})
\Sigma^{\lambda_1\lambda_2}({\bf k})G^{\lambda_2\beta}({\bf k}),
\end{equation}
with the self energy defined as,
\begin{equation}\label{eq:selfenergy}
  \Sigma^{\lambda_1\lambda_2}({\bf k})=TL^d
\sum_{n=1}^\infty \frac{(-1)^n}{n!}
\left\langle\left\|  \frac{\delta^2}{\delta u^{\lambda_1}_{-{\bf k}}
\delta u^{\lambda_2}_{{\bf k}}}
\left(
\frac{H'}{T}\right)^n
\right\|\right\rangle_{i,G_0},
\end{equation}
The subscript $i$ in the average tells us to only
include irreducible contributions, that is, any term which can be separated
by cutting just one leg should be ignored, as it is already included in the
full Green's function on the RHS of (\ref{eq:greens}).
A further resummation is possible by evaluating the RHS of
(\ref{eq:selfenergy}) with respect
to the renormalized, rather than the harmonic, Green's function. The
functional form of the self energy will then depend self-consistently on $G$,
\begin{eqnarray}\label{eq:selfenergyR}
  \Sigma^{\lambda_1\lambda_2}({\bf k},G)&&=\\
&&TL^d
\sum_{n=1}^\infty \frac{(-1)^n}{n!}
\left\langle\left\|  \frac{\delta^2}{\delta u^{\lambda_1}_{-{\bf k}}
\delta u^{\lambda_2}_{{\bf k}}}
\left(
\frac{H'}{T}\right)^n
\right\|\right\rangle_{s,G},\nonumber
\end{eqnarray}
where the subscript $s$ tells us only to include ``skeletal'' diagrams in the
average. We define a skeletal diagram as one that has not already been
included in the resummations of lower order diagrams
by using the full Green's functions. For instance the diagrams of group (c) in 
Fig.~\ref{fig:diagrams} must be ignored, as they are included when we evaluate 
the self-energy contribution from group (a) with the renormalized propagator.
After these
resummations we have a Dyson-type equation, which must be solved
self-consistently for $G$,
\begin{equation}\label{eq:dyson}
  (G^{-1})^{\alpha\beta}({\bf k})=(G_0^{-1})^{\alpha\beta}({\bf k})-
\Sigma^{\alpha\beta}({\bf k},G).
\end{equation}
Note that this inverse propagator is exactly the response function
$\Phi^{\alpha\beta}({\bf k})$ of the fluctuating crystal for an external
force. This is demonstrated in Appendix~\ref{ap:greens}.

\subsection{The (3+4)-SCA}

\begin{figure}
\centerline{\epsfxsize= 7.5cm\epsfbox{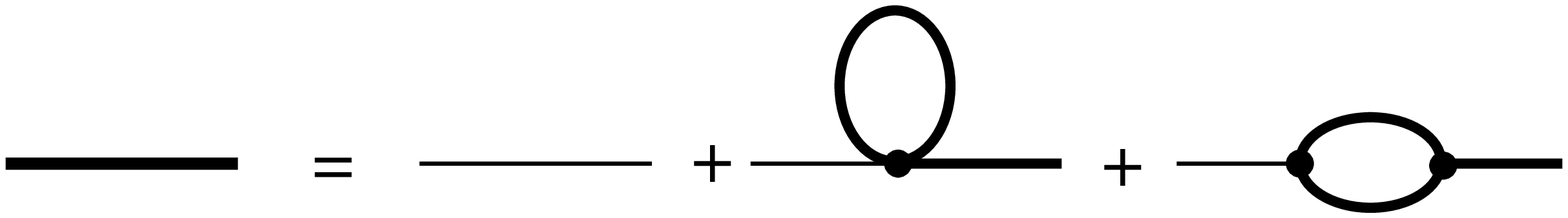}}
\vspace{0.2cm}
\caption{ The diagrammatic representation of the (3+4)-SCA. 
  The thick lines are the renormalized propagators. This is just the Dyson
  equation for the two lowest order diagrams, the ``Hartree'' diagram coming
  from $H_4$ and the ``flying-saucer'' diagram coming from $(H_3)^2$. }
\label{fig:diags_34}
\end{figure}

Due to the difficulties of calculating all possible diagrams in the
perturbation expansion, one looks for approximate resummations. A simple
example is to only take the lowest order skeletal diagrams (group (a) in
Fig.~\ref{fig:diagrams}) in the self energy, which is known as the
(3+4)-SCA.\cite{Lozovik} This is shown in
Fig.~\ref{fig:diags_34}, and it corresponds to solving the Dyson equation with
the self energy,
\begin{eqnarray}
\Sigma^{\alpha\beta}_{\rm 3+4}({\bf k},G)&=&
-L^d 
\left\langle \left\|
\frac{\delta^2 H_4}{\delta u^\alpha_{\bf k}
\delta u^\beta_{-{\bf k}}}
\right\|\right\rangle_{s,G}\nonumber\\
&&+\frac{L^d}{2T}
\left\langle\left\| 
\frac{\delta^2 ({H_3}^2)}{\delta u^\alpha_{\bf k}
\delta u^\beta_{-{\bf k}}}
\right\|\right\rangle_{s,G}.
\end{eqnarray}
Note that this method gives the exact propagator to order $T^2$, although it
only includes six of the thirteen diagrams at order $T^3$. The two skeletal
diagrams are typically of opposite sign and, importantly, for all the cases
of 2D power law interactions looked at by Lozovik {\em et al.} the magnitude
of the positive
${H_3}^2$ term is slightly greater than the negative $H_4$ term. This means the
lattice first softens with temperature, but any approximation that does not
include the ${H_3}^2$ term will incorrectly find a hardening of the lattice
with increasing temperature. More details of the (3+4)-SCA and calculations
for the example of logarithmic interactions in 2D are given in
Appendix~\ref{ap:34}.

\subsection{Diagrammatics of the SCHA}
\label{sec:diagsSCHA}

\begin{figure}
\vspace{-0.2cm}
\centerline{\epsfxsize= 8cm\epsfbox{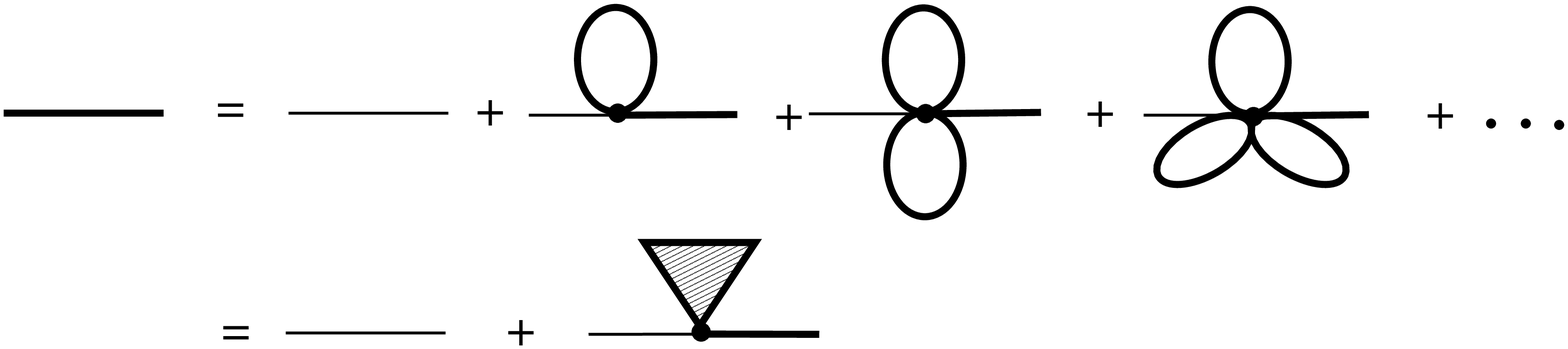}}
\vspace{0.2cm}
\caption{ The diagrammatic representation of the SCHA. The thick lines are the
  renormalized propagators. The filled triangle represents the self-energy in
the SCHA. The symmetry requirements of this approximation
  mean that only even vertices are included. }
\label{fig:diags_SCHA}
\end{figure}

It is insightful to consider which diagrams the SCHA (see
Section~\ref{sec:SCHA})  corresponds to. Writing $H=H_2+H'$ we see that
Eq.~(\ref{eq:eff}) becomes
\begin{equation}
  \Phi_t^{\alpha\beta}({\bf k})=
 \Phi_0^{\alpha\beta}({\bf k})
+L^d
\left\langle 
\frac{\delta^2 H'}{\delta u^\alpha_{\bf k}
\delta u^\beta_{-{\bf k}}}
\right\rangle_{t}.
\end{equation}
Comparing this to (\ref{eq:dyson}) and equating $\Phi_t$ to a Green's function
${G_t}^{-1}$
we see that the SCHA corresponds to solving
the Dyson equation with the self energy given by
\begin{equation}
\Sigma^{\alpha\beta}_{\rm SCHA}({\bf k},G_t)=
-L^d
\left\langle 
\frac{\delta^2 H'}{\delta u^\alpha_{\bf k}
\delta u^\beta_{-{\bf k}}}
\right\rangle_{t},
\end{equation}
which is equivalent to the $n=1$ part of
the sum in (\ref{eq:selfenergyR}) (for $n=1$ there are no
non-skeletal diagrams, so that the averages are equivalent). 
Therefore the SCHA is formally identical to
the diagrammatic equation in
Fig.~\ref{fig:diags_SCHA}.
Note that
only even order vertices are present, as the average of any odd combination
over the elastic Hamiltonian is zero. This is an important feature of the
SCHA. (An identical diagrammatic series appears for a form of SCHA
in the context of the sine-Gordon model\cite{Minnhagen}. In this case, the
cosine potential only has terms even in the field, and so the 
SCHA has a better chance of being an accurate approximation.)
For our problem the SCHA only includes one of the two diagrams that
contribute to the propagator at order $T^2$, and only three of the thirteen
diagrams at the next order (compare to Fig.~\ref{fig:diagrams}). 
The usefulness of the SCHA is that it is straightforward to calculate averages
of the true Hamiltonian over a Gaussian distribution. This avoids the
difficult task of actually evaluating high order vertices in a diagrammatic 
expansion. (E.g. for a pairwise interaction one must evaluate a sum over 
$m$ derivatives of the potential to find the $m$-th vertex. This is done for
$m=3$ and $m=4$ in Appendix~\ref{ap:34}.)

\section{A new self-consistent method}\label{sec:new}

\begin{figure}
\centerline{\epsfxsize= 8cm\epsfbox{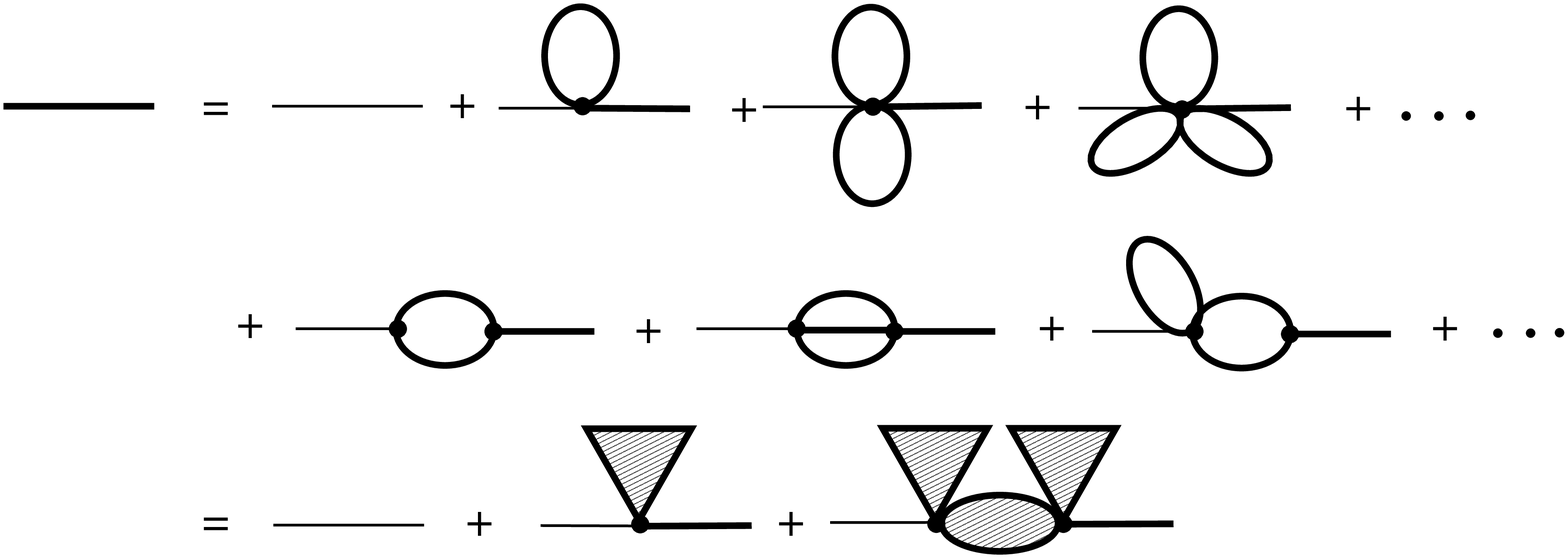}}
\vspace{0.2cm}
\caption{ The two-vertex-SCHA described in diagrams. This
  approximation includes nine of the thirteen diagrams to order $T^3$ in
  Fig.~\ref{fig:diagrams}. This is equivalent to the self-consistent solution
  of Eq.~(\ref{eq:neweff}), so that the diagrams do not need to be explicitly
  calculated.}
\label{fig:diags_n2}
\end{figure}

We want to develop a method that combines the simplicity and elegance of the
SCHA with the accuracy at lowest order of the (3+4)-SCA. We note that the SCHA
is equivalent to evaluating just the $n=1$ terms in the self energy 
(\ref{eq:selfenergy}),
while it is  the $n=2$ part that
includes the ``flying-saucer'' diagram that is so important in the
(3+4)-SCA. The contribution to the self energy from $n=2$ is,
\begin{equation}
\Sigma_{n=2}^{\lambda_1\lambda_2}({\bf k},G)=
\frac{L^d}{2T}
\left\langle\left\|  \frac{\delta^2}{\delta u^{\lambda_1}_{-{\bf k}}
\delta u^{\lambda_2}_{{\bf k}}
}
\left(H'\right)^2
\right\|\right\rangle_{s,G}.
\end{equation}
To write this as an average over the full Hamiltonian with a given
distribution of displacements,
we must subtract the diagrams that were
included in resummations of the $n=1$ diagrams. Allowing for this, and 
separating $H'=H-H_2$ leads to the form,
\begin{eqnarray}
\Sigma_{n=2}^{\lambda_1\lambda_2}({\bf k},G_t)&=&
\frac{L^d}{T}
\left\langle  \frac{\delta H}{\delta u^{\lambda_1}_{-{\bf k}}}
 \frac{\delta H}{
\delta u^{\lambda_2}_{{\bf k}} }
\right\rangle_t\\
&&\hspace{-2cm}-\frac{L^d}{T}
\left\langle  u^{\lambda_3}_{-{\bf k}} u^{\lambda_4}_{{\bf k}}
\right\rangle_t
\left\langle   \frac{\delta^2 H}{\delta u^{\lambda_1}_{-{\bf k}}
\delta u^{\lambda_3}_{{\bf k}}}
\right\rangle_t
\left\langle   \frac{\delta^2 H}{\delta u^{\lambda_4}_{-{\bf k}}
\delta u^{\lambda_2}_{{\bf k}}}
\right\rangle_t.\nonumber
\end{eqnarray}
Note that terms explicitly including $H_2$ have cancelled with each other.
This corresponds to calculating, within an SCHA type method, the effective
elastic matrix,
\begin{eqnarray}\label{eq:neweff}
\Phi_t^{\alpha\beta}({\bf k})&=& L^d
\left\langle  \frac{\delta^2 H}{\delta u^{\alpha}_{-{\bf k}}
\delta u^{\beta}_{{\bf k}}}
\right\rangle_t\nonumber\\
&&-\frac{L^d}{T}\left[
\left\langle  \frac{\delta H}{\delta u^{\alpha}_{-{\bf k}}}
 \frac{\delta H}{
\delta u^{\beta}_{{\bf k}} }
\right\rangle_t\right.\\
&&\hspace{-2cm}\left. -
\left\langle  u^{\lambda_1}_{-{\bf k}} u^{\lambda_2}_{{\bf k}}
\right\rangle_t
\left\langle   \frac{\delta^2 H}{\delta u^{\alpha}_{-{\bf k}}
\delta u^{\lambda_1}_{{\bf k}}}
\right\rangle_t
\left\langle   \frac{\delta^2 H}{\delta u^{\lambda_2}_{-{\bf k}}
\delta u^{\beta}_{{\bf k}}}
\right\rangle_t
\right].\nonumber
\end{eqnarray}
We call this equation the ``two-vertex-SCHA''. 
The extra terms look rather like a fluctuation contribution to the elasticity,
but we should remember that it is simply the two-vertex part of the Dyson
series, and there will in principle be further corrections with higher numbers
of vertices. The diagrammatic representation of the two-vertex-SCHA is
shown in Fig.~\ref{fig:diags_n2}.
This scheme now incorporates the diagrams from the SCHA and the (3+4)-SCA
along with more terms not included in these last two approaches. For instance,
it includes nine of the thirteen
diagrams for the propagator at order $T^3$, see Fig.~\ref{fig:diagrams}. 
It also has the feature of the
SCHA that one only needs to calculate averages of derivatives of
the full Hamiltonian, and  not any diagrams explicitly.

\section{Physical realizations} \label{sec:real}

\subsection{The 2D Coulomb system}\label{sec:2D}

The 2D Coulomb system, also called the one-component plasma, consists of
particles interacting with a pairwise, repulsive logarithmic potential,
\begin{equation}\label{eq:logint}
  V(R)=-v\ln(R/\xi).
\end{equation}
It has been studied in the past partly because of some special mathematical
properties\cite{Alastuey_Jancovici} 
(the system is exactly soluble at one particular temperature $T=v/2$ in the 
liquid phase).  Numerical 
simulations\cite{deLeeuw_Perram,Caillol,Choquard_Clerouin}
have found evidence for a weak first order
transition at $T_m\approx 0.007v$.
Physically, this system is of interest as it is applicable to a thin
superconducting film with field-induced vortices which interact 
logarithmically\cite{Pearl} up to 
a screening length $\Lambda=\lambda^2/d$ where $d$ is the thickness of the
film and $\lambda$ is the bulk penetration depth, typically of order 1000~\AA.
Such films were predicted to undergo a 2D melting transition of
the vortex lattice well below the zero-field transition 
temperature.\cite{Huberman,Fisher} (A similar scenario takes place in layered
superconductors at high fields, see Ref.~\onlinecite{review}.)

The ground state of the 2D
Coulomb model is a triangular lattice, with nearest-neighbor spacing 
$a=(2/n\sqrt{3})^{1/2}$. This lattice
has a shear modulus of 
\begin{equation}
\mu_0=\lim_{k_y\rightarrow 0}\left[\Phi^{xx}_0(k_y)/{k_y}^2\right]=
\frac{nv}{8}
\end{equation}
(this result was known over thirty years ago\cite{FHP}) and a 
diverging compression modulus at large wavelengths, 
\begin{equation}
\lambda_0(k)=\Phi^{xx}_0(k_x)/{k_x}^2
\approx 2\pi vn^2/k^2.
\end{equation}
In Refs.~\onlinecite{Huberman} and~\onlinecite{Fisher} it was presumed that 
the melting
transition would be of the continuous
Kosterlitz-Thouless-Halperin-Nelson-Young (KTHNY) type\cite{KT,HN} 
with a dislocation-unbinding mechanism, which is predicted to occur at
\begin{equation}
  T_{\rm KTHNY}=\mu\frac{a^2}{4\pi}=A(T) \frac{v}{16\sqrt{3}\pi}
\approx 0.011 A(T)v
\end{equation}
with $A<1$
 the renormalization constant for the shear modulus at finite
temperature, $\mu=\mu_0A(T)$. 
However, this is not consistent
with the first-order result of 
simulations.\cite{deLeeuw_Perram,Caillol,Choquard_Clerouin}
One possible reason for this discrepancy is that regular fluctuations in an
anharmonic potential, such as 
we considered in the last section, can
destabilize the lattice before singular 
fluctuations such as free dislocations are generated.
Therefore in this section we consider 
the anharmonic softening and instability of the 2D Coulomb lattice, and
compare to the predictions of the KTHNY theory. We calculate
$\mu$ within the SCHA. We also find the first order correction
in temperature
to the 
shear modulus using perturbation theory, by calculating the 
diagrams of group (a) in Fig.~\ref{fig:diagrams}.
This shows that the SCHA greatly
underestimates the thermal softening. We then present our results for $\mu$
from the (3+4)-SCA and the two-vertex SCHA and compare to the KTHNY 
universal value of $\mu=4\pi T_{\rm KTHNY}/a^2$.

The SCHA for the 2D Coulomb system was briefly looked at in 
Ref.~\onlinecite{Alastuey_Jancovici}. In fact,
the SCHA with ``independent fluctuations''  has some nice features for this
system due to the Laplacian of the interaction being zero. In 
Appendix~\ref{ap:log} 
we derive the non-perturbative form of the elastic moduli as a function of
$\sigma_{\mu 0}=\frac 12\langle \left| {\bf u}({\bf R}_\mu)-
{\bf u}(0)\right|^2\rangle$, 
which leads to the result for the shear modulus,
\begin{eqnarray}\label{eq:c66ofu2}
 \mu_t &=&\mu_0\sum_{{\bf R}_\mu}
\left(1-\frac{{R_\mu}^2}{2\sigma_{\mu0}}\right)
e^{-{R_\mu}^2/2\sigma_{\mu0}}\\
&\approx &
\mu_0+ 6\mu_0
\left(1-\frac{a^2}{2\langle u^2\rangle_{\rm nn}}\right)
e^{-a^2/2\langle u^2\rangle_{\rm nn}}.
\end{eqnarray}
This formula is shown in the inset of Fig.~\ref{fig:2D_c66ofu2}
(the nearest-neighbor approximation here is extremely accurate). 
The non-perturbative feature (the extreme
flatness at small $\langle u^2\rangle_{\rm nn}$) will be lost
when we include the correlations in fluctuations between different particles.
In fact, we have found that the SCHA predicts an initial stiffening of lattice with
small fluctuations, which we understand to be an incorrect result due to 
the neglect of diagrams with odd vertices.

\begin{figure}
\centerline{\epsfxsize= 8cm\epsfbox{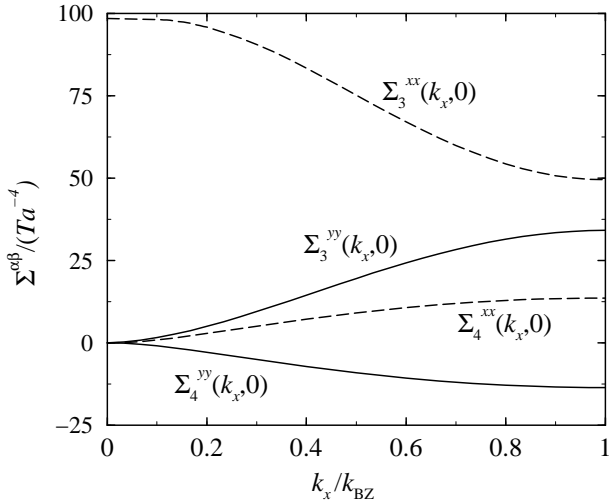}}
\caption{ Contributions to the transverse and longitudinal self energy
from the skeletal diagrams of Fig.~\ref{fig:diags_34}, as given by
Eqs.~(\ref{eq:flyingsaucer}) and~(\ref{eq:hartree}), for logarithmic pairwise
interactions in 2D.
}
\label{fig:S3S4_2D}
\end{figure}

Inspection of Eq.~(\ref{eq:dyson}) shows the correct temperature expansion of
the change in shear modulus to be,
\begin{equation}\label{eq:deltamu}
 \mu=\mu_0 -\lim_{k\rightarrow 0}\Sigma^{xx}(k_y,G)/{k_y}^2. 
\end{equation}
We can find the first correction to the shear modulus to order $T$ by
evaluating the bare self-energy contributions in 
the diagrams of Fig.~\ref{fig:diags_34}. Formulas for the (3+4)
self-energies for a system with pairwise interactions are given in
Appendix~\ref{ap:34}. We have numerically evaluated these formulas
for the 2D Coulomb system, and our results for the transverse 
$\Sigma^T=\Sigma^{yy}(k_x)$
and longitudinal $\Sigma^L=\Sigma^{xx}(k_x)$
parts taking $G=G_0$ are shown in Fig.~\ref{fig:S3S4_2D}. 
The large wavelength limits are found to go as,
\begin{equation}\label{eq:selftrans}
\Sigma_3^T(k)\approx (14 T/a^2)k^2\,\,\,\,\,\,\,\,\,
\Sigma_4^T(k)\approx-(10 T/a^2)k^2
\end{equation}
\begin{equation}\label{eq:selflong}
\Sigma_3^L(k)\approx 98 T/a^4
\,\,\,\,\,\,\,\,\,
\Sigma_4^L(k)\approx(10 T/a^2)k^2
\end{equation}
From Eqs.~(\ref{eq:deltamu}) and~(\ref{eq:selftrans}) 
the shear modulus at low temperatures will be
given by 
\begin{equation}\label{eq:mufirst}
\mu\approx\mu_0\left(1-28\frac{T}{v}\right),
\end{equation} 
so that within first-order perturbation theory, $\mu$ is about $80\%$ of its
zero temperature value at the numerically observed transition at 
$T_m\approx 0.007v$.
The large
wavelength limit of the longitudinal part of the self energy in
(\ref{eq:selflong}) is a constant, so 
that the compressional modes are also softened to first order in temperature,
\begin{equation}
\lambda(k)\approx\frac{2\pi n^2v}{k^2} 
\left(1 - 12\frac{T}{v}\right).
\end{equation} 
This thermal softening of the long-wavelength compression energy (a special 
feature of the 2D Coulomb system) might seem to contradict the physically 
intuitive expectation that the compressibility is fixed by the bare charge of 
the system.
A full understanding of this result will be discussed in a future paper, but 
it is due to the non-linear relation between changes in the density and the 
longitudinal displacements, 
$\delta n = n \nabla\cdot {\bf u} + {\cal O}\left(u^2\right)$.

\begin{figure}
\centerline{\epsfxsize= 8cm\epsfbox{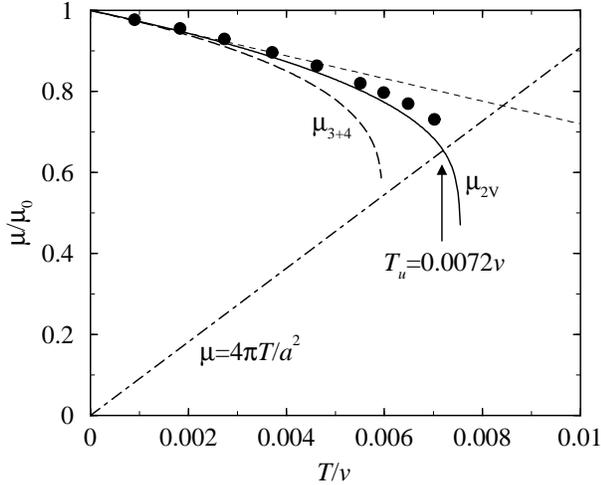}}
\caption{ The shear modulus of the 2D Coulomb crystal as a function of
  temperature as calculated to  first-order in Eq.~(\ref{eq:mufirst}) 
  (short dashed
  line), and within the (3+4)-SCA (long dashed line) and the 
  two-vertex-SCHA (full
  line). At higher orders in $T$ we see that the two-vertex-SCHA predicts less 
  softening than the (3+4)-SCA, and a correspondingly higher instability 
  temperature, $T_u$. The circles are the measured values of $\mu$ from
  numerical simulations taken from Ref.~36. 
  They agree well with the two-vertex-SCHA 
  curve. The dot-dashed line follows the universal jump
  prediction of the KTHNY transition, $\mu=4\pi T/a^2$. Below this line the 
  lattice is unstable to free dislocations. }
\label{fig:c66oft_2d}
\end{figure}

We can go beyond the lowest order in $T$ using the (3+4) self energy by
iterating the equation 
of Fig.~\ref{fig:diags_34} until a stable solution is found within this
(3+4)-SCA.
We have also found numerical solutions within the two-vertex-SCHA
of Section~\ref{sec:new}. Details for calculating the effective elastic moduli 
of Eq.~(\ref{eq:neweff}) 
for pairwise interactions are given in Appendix~\ref{ap:n=2}. As we
have shown in Section~\ref{sec:new}, this approximation gives identical
results to lowest order in $T$ as the (3+4)-SCA. However, we find important
deviations at larger temperatures.
In Fig.~\ref{fig:c66oft_2d} the results for $\mu$ as a function of temperature
from both methods are shown. 
The shear modulus was recently measured in the same system with 
Langevin-dynamics simulations,\cite{Cai} and we find good agreement
between
the measured temperature dependence (the circles in Fig.~\ref{fig:c66oft_2d})
and the two-vertex-SCHA calculation.  
Notice that the instability point from the  (3+4)-SCA is below
the numerically measured melting temperature of $T_m\approx 0.007 v$, 
while the 
upper limit to solutions of the two-vertex-SCHA is slightly above this
temperature. 

In Fig.~\ref{fig:c66oft_2d} we also 
compare to the predictions of the KTHNY theory 
of 2D melting. As described above, this theory predicts a universal value of
the shear modulus at the dislocation-unbinding transition, 
$\mu=4\pi T/a^2$, shown as the dot-dashed line in Fig.~\ref{fig:c66oft_2d}. 
The fact that the two-vertex-SCHA result crosses this
value tells us that the results at temperatures above this value are
unphysical, as the crystal is unstable to free dislocations (if the (3+4)-SCA
were correct, it would practically rule out the KTHNY mechanism for this
system). Therefore the crossing of
the universal jump and the  two-vertex-SCHA result is an upper bound to a
melting transition and we find $T_u= 0.0072 v$ 
(the effect of thermally generated dislocation pairs will
push $\mu$, and the melting temperature, down further).
The fact that at melting the shear modulus is so close to the 
KTHNY universal
value tells us that dislocation-pairs (not included in our analysis) as well
as anharmonicities both play an important role in the physics of the
melting transition. Whether the weak first-order nature seen in simulations is 
correct, or a numerical artifact, remains unclear.

\subsection{Classical electrons in two dimensions}\label{sec:Wigner}

It is experimentally feasible to confine electrons to a flat surface
in various systems, such that they are free to move in only two dimensions. In 
particular, for electrons trapped on the surface of liquid helium by an
applied electric potential, a classical melting transition from a 2D ``Wigner
crystal'' to a liquid state has been observed.\cite{GrimesAdams}
The elastic properties of the 2D Wigner crystal are well
studied,\cite{Crandall,Bonsall} 
and Platzman and Fukuyama\cite{PlatzmanFukuyama} considered the effect 
of thermal fluctuations using the SCHA, reporting a dramatic stiffening of the 
lattice with temperature. Using the perturbation expansion of
Section~\ref{sec:pert}, Fisher\cite{Fisher2}
 showed this to be incorrect, as the first order
term in $T$ gives a downward correction to $\mu$.
Lozovik {\em et al.} have applied their (3+4)-SCA to this
problem,\cite{Lozovik} and stated that an anharmonic instability is reached
below the KTHNY temperature.
However, the discrepancy in Section~\ref{sec:2D} for logarithmic interactions 
between the  (3+4)-SCA  and
our two-vertex-SCHA  motivates us to reconsider
the 2D Wigner crystal.

Electrons interact with a repulsive pairwise potential,
\begin{equation}
  V(R)=\frac{e^2}R,
\end{equation}
where $e$ is the electron's charge. The ground state is a triangular lattice,
spacing $a$, and at finite temperatures all properties depend on the
dimensionless ratio $T/(e^2a^{-1})$. In the literature the parameter 
$\Gamma=\sqrt{\pi}n^{\frac 12}e^2/T$ is more often used. When calculating the
elastic matrix, care must be taken to split the interaction to
long- and short-range parts, and treat them separately as in the Ewald
technique.\cite{Bonsall} Then the zero-temperature 
shear modulus is found to be,
\begin{equation}
  \mu_0=0.2450645\,\, e^2n^{3/2},
\end{equation}
and the compression modulus diverges at small $k$ (as for the Coulomb crystal
but with a different power) as
\begin{equation}
  \lambda_0(k)=\frac{2\pi e^2n^2}{k}.
\end{equation}
Repeating the perturbation theory of Section~\ref{sec:pert} we find the
first-order correction to the shear modulus to be,
\begin{equation}
  \mu\approx \mu_0\left(1-15 \frac{T}{e^2 n^{1/2}}\right),
\end{equation}
in agreement with Ref.~\onlinecite{Fisher2}. This correction again comes from a
negative contribution from $\Sigma^T_3(G_0)$ and a positive, but smaller in
magnitude, contribution from $\Sigma^T_4(G_0)$.
We have also confirmed that the SCHA gives an incorrect stiffening of $\mu$
with increasing temperature.

\begin{figure}
\centerline{\epsfxsize= 8cm\epsfbox{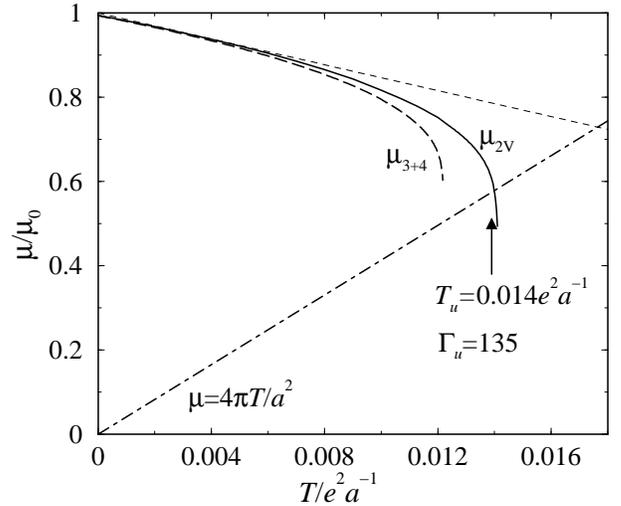}}
\caption{ The shear modulus of the 2D Wigner crystal as a function of
  temperature. We show the first-order correction of Ref.~8
  (short dashed
  line), and our results for the (3+4)-SCA (long dashed line) and the 
  two-vertex-SCHA (full  line). 
  The results are qualitatively the same as for the Coulomb crystal in 
  Fig.~\ref{fig:c66oft_2d}
  with the two-vertex-SCHA result crossing the KTHNY universal jump line
  (dot-dashed line).
 }
\label{fig:c66oft_wig}
\end{figure}

In Fig.~\ref{fig:c66oft_wig}
 we plot our results for the shear modulus $\mu(T)$ from the
(3+4)-SCA and the two-vertex-SCHA. 
The resulting picture is very similar to that for logarithmically interacting
particles in Fig.~\ref{fig:c66oft_2d}.
Our instability point for the 
(3+4)-SCA of $\Gamma^{\rm 3+4}_u\approx 155$ 
is close to the quoted value from Lozovik
{\em et al.} of $\Gamma^{\rm 3+4}_u=140$. The two-vertex-SCHA is again more
stable, and together with the KTHNY condition on the minimum allowed  $\mu(T)$
gives $\Gamma^{\rm TV}_u\approx135$. 
This is close to  the experimental 
melting temperature of $\Gamma_m=127\pm 5$,\cite{GrimesAdams,Deville} valid
over a range of experimental densities. Also,
numerical simulations have found a melting transition in the range from
$\Gamma^{\rm sim}_m\approx90$ (Ref.~\onlinecite{Hockney}) to  
$\Gamma^{\rm sim}_m\approx125$ 
(Ref.~\onlinecite{Gann} and~\onlinecite{Morf}) to 
$\Gamma^{\rm sim}_m\approx159$ (Ref.~\onlinecite{Bedanov}). 
(It is not clear why
there is such disagreement between the numerical results. The rather high
melting temperature from the earliest
simulation\cite{Hockney} is probably an out of equilibrium artifact; the
system was started in the crystal state and heated up.)

\subsection{Quantum melting of 2D bosons 
and the Bose model of vortex-lattice melting}\label{sec:VL}

In the previous sections we have ignored the effect of quantum
fluctuations. The uncertainty principle can prevent the formation of a lattice 
phase if the mass of the particles is sufficiently low. In this section we
will consider the effect of quantum fluctuations on a 2D Coulomb lattice of
bosons at zero temperature. In the path-integral representation, the action of 
the boson system in imaginary time is,
\begin{equation}
  S=\int^{\hbar/T}_0 d\tau\,\,\frac m 2 \sum_\mu 
\left|\frac{d{\bf R}_\mu}{d\tau}\right|^2
+\frac 1 2\sum_{\mu\ne\nu} V( \left|{\bf R}_\mu-{\bf R}_\nu\right|),
\end{equation}
with $V(R)$ given in Eq.~(\ref{eq:logint}).
The properties of the system are determined by the statistical sum, or
partition function, $Z=\hbox{Tr}_{{\bf R}_\mu(\tau)}\left[e^{-S/\hbar}\right]$. 
At zero temperature all properties depend only on the unitless de~Boer 
parameter,
$\Lambda=\left(\hbar^2/a^2vm\right)^{1/2}$,
which measures the ratio of zero-point kinetic energy to the interaction
energy. 

This model maps directly to a 
3D classical system at finite temperature, where the world lines of the bosons 
correspond to fluctuating elastic strings. In fact, the Coulomb interaction
between the strings provides a model of the interactions between vortex lines
in a type-II superconductor (this fruitful analogy was discovered by
Nelson\cite{Nelson}). 
Thus the path-integral representation of 2D bosons has been studied
with the motivation of describing vortex-lattice
melting\cite{Nordborg,Nordborg2}  which 
occurs at finite magnetic fields in high-$T_c$ superconductors.\cite{review}
To be specific, the 3D vortex system is
obtained by replacing
$\hbar$ with $T$, the time $\tau$ with the distance along the field
direction $z$ 
and the mass $m$ with the line elasticity 
$\varepsilon_l$. (In fact, this model only approximates the 3D interactions of
the vortex system. A more correct mapping would be from a boson system with 
interactions non-local in time.) 
The strength of the interaction per unit length
is determined by the
penetration depth $\lambda$ of the superconductor via the relation
$v\rightarrow 2\varepsilon_0$ with $\varepsilon_0=(\phi_0/4\pi\lambda)^2$
where the quantum of flux is $\phi_0=hc/2e$.
In these units the de~Boer parameter is
\begin{equation}
\Lambda=T\left(a^22\varepsilon_0\varepsilon_l\right)^{-1/2}.
\end{equation}
In the numerical simulations\cite{Nordborg,Nordborg2} a melting transition
occurs at $\Lambda\approx 0.062$, above which the system is in a line-liquid
phase. Although it has proved difficult to measure the equilibrium values of
the elastic moduli in the same simulations, the value of $\langle u^2\rangle$
was measured, and shows deviations from a linear dependence on temperature
before melting occurs.\cite{Nordborg2} In this section we use our
self-consistent treatment to find the effect of anharmonicity and the 
instability point in this system and compare our results to the measured 
$\langle u^2\rangle$ in the simulations. We also
calculate the quantitative effect of fluctuations on the shear modulus of the
vortex crystal.

In the following we use the notation for the classical vortex system, but all
results are easily transferred to the 2D quantum system at zero
temperature. The generalizations to fluctuating lines of the results of
Sections~2--5 are straightforward, where the displacement variables are 2D
vectors in 3D ${\bf k}$-space,
\begin{equation}
 {\bf u}_{\bf k}=\frac{1}{n}
 \int dz \sum_{{\bf R}_\mu}{\bf u}_\mu(z) 
e^{-i({\bf k}_\perp.{\bf R}_\mu+k_zz)}.
\end{equation}
The elastic matrix and the self energy are now functions of $k_z$ as well as
the 2D vector ${\bf k}_\perp$. Brief details of the (3+4)-SCA and the
two-vertex-SCHA are given in Appendix~\ref{ap:string}.

At large wavelengths the transverse elastic matrix is given by
$\lim_{k\rightarrow 0}\Phi^T({\bf k})= c_{66}k_\perp^2+c_{44}k_z^2$ and the
longitudinal modes by 
$\lim_{k\rightarrow 0}\Phi^L({\bf k})= c_{11}k_\perp^2+c_{44}k_z^2$
where $c_{11}$, $c_{66}$ and $c_{44}$ are the Voigt notation for the
compression, shear and tilt moduli respectively. 
We can again find the first correction to the shear modulus by evaluating
the self energy to lowest order in $T$. We find that
$\Sigma_3^T(k_\perp)\approx 3.2 (T/a^2a_z)
k_\perp^2$ and
$\Sigma_4^T(k_\perp)\approx-2.0  (T/a^2a_z)
k_\perp^2$,
where $a_z=a
\left(\varepsilon_l/\varepsilon_0\right)^{\frac 12}$.
Therefore to 
lowest order in the de~Boer parameter,
\begin{equation}\label{eq:bosec66_1st}
c_{66} \approx c_{66}^0\left(1-5.5\Lambda\right),
\end{equation}
where the shear modulus without fluctuations is $c_{66}^0=n\varepsilon_0/4$.
The compression modulus at large wavelengths is not renormalized.
Note that the
``Hartree'' self-energy $\Sigma_4({\bf k})$ is independent of the
$z$-component of the wave-vector 
(see Appendix~\ref{ap:string}), but the contribution $\Sigma_3({\bf k})$ does
depend on $k_z$. For example, for ${\bf k}_\perp$ at the Brillouin zone we
find,
$\Sigma_3^{xx}(k^x_\perp=k_{\rm BZ},k_z)\approx
(T/a^4a_z)(8.9
-2.8 a_z^2k_z^2)$, which gives a fluctuation induced stiffening of the tilt
modulus, 
\begin{equation}
  \label{eq:c443d}
  c_{44}(k^x_\perp=k_{\rm BZ})\approx \frac{\varepsilon_l}{a^2}\left(
1+3.9\Lambda\right).
\end{equation}

We have made detailed calculations for the two-vertex-SCHA for this model, and 
the results for the shear modulus as a
function of $\Lambda$
are shown in Fig.~\ref{fig:c66oft_3d}.
It turns out that the effective elastic matrix (see Appendix~\ref{ap:string})
depends on the full correlations of displacement fluctuations of a given
vortex line in the $z$-direction, and we need to numerically find a
self-consistent solution for the function,
\begin{eqnarray}
  \label{eq:u2ofz}
\langle u^2(z)\rangle &\equiv& \langle u^\alpha(z)u^\alpha(0)\rangle\\
&=& T \int^\infty_{-\infty} \frac{dk_z}{2\pi}
 \int_{\rm BZ} \frac{d^2k_\perp}{(2\pi)^2}
e^{-ik_zz} (\Phi_t^{-1})^{\alpha\alpha}({\bf k}).\nonumber
\end{eqnarray}
In practise we discretize along the $z$-direction, making sure that
$\langle u^2(z)\rangle$ is smoothly varying.
\begin{figure}
\centerline{\epsfxsize= 8cm\epsfbox{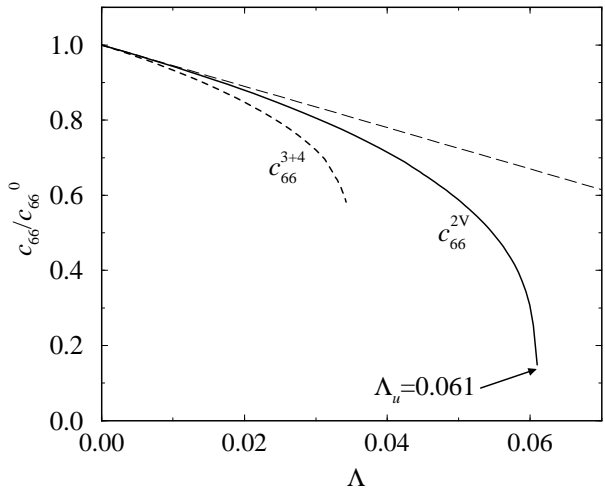}}
\caption{ Shear modulus for 2D bosons with quantum fluctuations, or for the
  Bose model of the vortex lattice at finite temperature.
The long-dashed line is the first-order result (\ref{eq:bosec66_1st}). The
short-dashed line is the result from the (3+4)-SCA, while the full line is the 
shear modulus calculated in the two-vertex-SCHA. Note that a stable solution
is found for much smaller values of the shear modulus than for a
two-dimensional system at finite temperature (see Fig.~\ref{fig:c66oft_2d}).
}
\label{fig:c66oft_3d}
\end{figure}

 We again see that the stability limit 
from the two-vertex-SCHA is higher than that from the (3+4)-SCA. Surprisingly, 
the crystal maintains a self-consistent solution for a shear modulus which is
only 20\% of the zero-temperature value. 
This contrasts with our results for a 2D
crystal, and is mainly due to the stiffness of the elastic lines: the average
displacements within the equipartition theorem only grow with the inverse
square-root of the shear modulus,\cite{review} 
$\langle u^2\rangle\propto T/\sqrt{c_{66}c_{44}}$. 
The result is qualitatively similar to the measured shear modulus in
simulations of coupled
vortex pancakes in a layered superconductor in Ref.~\onlinecite{Cai}. In their 
somewhat different model of a vortex lattice, a reduction in the shear modulus 
at melting of about 30\% of the zero temperature value was found. 
Also, we find that the tilt
modulus stiffens slightly with the fluctuations, allowing for a still softer 
shear modulus before instability is reached.

\begin{figure}
\centerline{\epsfxsize= 8cm\epsfbox{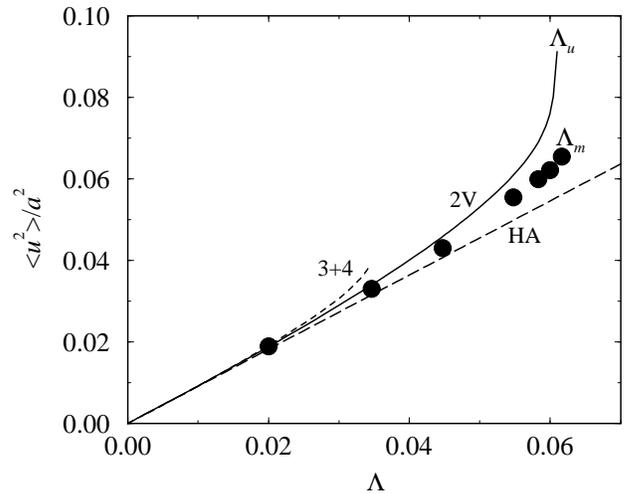}}
\caption{ Comparison of the single-particle fluctuations as measured in Monte
  Carlo simulations of Ref.~46
 and as calculated in this paper. The circles 
  are data from Ref.~46 and the full line is from the two-vertex SCHA, with a
  stability limit close to the numerical melting. The long dashed line
  is the harmonic result. The (3+4)-SCA (short-dashed line) 
  becomes unstable well below melting, while the instability 
  point of the SCHA (not shown) overestimates the melting point by a 
  factor two.}
\label{fig:u2oft_3d}
\end{figure}
In Fig.~\ref{fig:u2oft_3d}
we compare the mean fluctuation per particle 
$\langle u^2(0)\rangle=(2\pi)^{-3}\int d^3k \langle u^\alpha_{\bf k}
u^\alpha_{-{\bf k}}\rangle$
with the values measured in simulations
(see Fig.~8 of Ref.\onlinecite{Nordborg2}). The simulation results come from
fitting a Debye-Waller factor to the measured widths of the crystal structure
factor.\cite{Nordborg2} The main point is that the instability point predicted 
by the two-vertex-SCHA, $\Lambda_u=0.061$ is very close the numerically 
measured melting point $\Lambda_m=0.062$. The fact that we find $\Lambda_u$
slightly below melting means that the two-vertex-SCHA still slightly
overestimates the level of softening, and correspondingly there is an
increasing discrepancy with the numerical results for $\langle u^2\rangle$
close to the melting temperature. We should of course remember that the
two-vertex-SCHA is only a limited resummation of the perturbation
theory. However, our new method clearly gives much superior results compared
to the (3+4)-SCA, also shown in the figures, and to the SCHA, which we found
to give very little change to the elastic moduli below the numerically found
melting and an instability bound over twice the value of $\Lambda_m$.

The thermal softening of the shear modulus of the vortex
lattice is rarely explicitly calculated in the literature. 
The measurements from simulations in Ref.~\onlinecite{Cai} agree with 
Fig.~\ref{fig:c66oft_3d} in that
this softening is really significant. It may help to  explain the peak
effect in the critical current
seen close to the vortex-lattice melting transition in YBCO.\cite{PeakEffect}
The weak collective pinning theory predicts that, 
for a non-dispersive tilt modulus, the critical current depends on the elastic 
moduli as $j_c\propto\gamma^2/c_{66}^2c_{44}$ where $\gamma$ is a pinning
strength parameter.\cite{LarkinOvchinnikov}
However, the effect of finite temperature is not only to reduce the shear
modulus, but also to smooth the effective pinning landscape over $\langle
u^2\rangle$, thus reducing the pinning force. To see if our
results lead to a critical-current peak as a function of temperature requires
a careful analysis of the ${\bf k}$ dependent elastic moduli (the critical
pinning volume depends on long-wavelength distortions but  $\langle
u^2\rangle$ comes from integrating over all modes), and we leave this for a
future project.

\section{Conclusions}

In this paper we have introduced a new method, the two-vertex-SCHA, to treat
the anharmonicity of simple lattices in the presence of thermal, or quantum,
fluctuations. The method is based on the diagrammatic expansion about the HA,
but takes the form of an extension to the SCHA calculation of effective 
elastic moduli.
By considering two long-range interacting systems at finite
temperature, and a system of zero-temperature 2D bosons, we have seen that
the two-vertex-SCHA is a significant improvement over
previously used self-consistent techniques, the SCHA and the (3+4)-SCA, which
respectively under- and overestimated the fluctuation-induced softening.
It is the comparison to
simulation results that gives us the most confidence in the reliability of the 
two-vertex-SCHA as a tool for calculating the softening and instability of
such simple crystals due to thermal or quantum fluctuations.
One should therefore not trust the results of previous treatments of the
anharmonic softening of 2D lattices,\cite{PlatzmanFukuyama,Lozovik} 
but use instead the two-vertex-SCHA.

In addition to establishing this new method, we have found important results
for the vortex-lattice melting problem, using the Bose model for
the finite-temperature vortex system. We have found a much larger than
expected thermal softening of vortex lattice with the shear modulus reducing
to 20\% of its zero temperature value at the melting transition, which may
have
important consequences for the pinning properties of the vortex lattice.
It should be noted, however, that the real vortex system does not only have
interactions within planes of equal height, as in the Bose model. The full
London model of vortices corresponds to a 2D Bose system with 
time-nonlocal interactions.\cite{Feigelman} An investigation to extend the
results of the Bose model to the full time-nonlocal case is 
underway.\cite{Fominov}
Another facet of the vortex system in high-$T_c$ superconductors is the
layered nature of the cuprates. For very weak inter-layer coupling, the system 
can be described by magnetically coupled ``pancake'' vortices. 
In a recent paper the two-vertex-SCHA has been used to find the instability 
line of such a pancake-vortex lattice.\cite{Pancakes} 
It lies just above the melting line 
as calculated with a free energy comparison using results from 2D 
simulations for the pancake-vortex liquid phase.
Finally, it would also be interesting to investigate the stability of lattices 
in the presence of quenched disorder. This is also relevant for the vortex
lattice, where pinning disorder exists, and a
disorder-induced melting transition occurs
as the vortex density, or magnetic field,
is increased.\cite{Disorder}

\acknowledgements

We thank Anne van Otterlo, Henrik Nordborg, Alex Koshelev,
Boris Ivlev and Valerii Vinokur for useful discussions, and the Swiss 
National Foundation for financial support. MJWD is supported by an 
EPSRC Advanced Fellowship AF/99/0725.

\end{multicols}
\widetext

\appendix
\section{Nearest-neighbor fluctuations in two dimensions}
\label{ap:nn}

In the SCHA defined by Eq.~(\ref{eq:effpair}), 
we have to take averages over the second 
derivative of the potential between two particles as they fluctuate within 
the crystal (similar averages are needed in the two-vertex-SCHA, see
Appendix~\ref{ap:n=2}). Within the 
Gaussian approximation these averages only depend on the mean square,
$\sigma_{\mu 0}=\frac 12\langle \left| {\bf u}({\bf R}_\mu)-
{\bf u}(0)\right|^2\rangle$, which in two dimensions is given by
\begin{equation}
\sigma_{\mu 0}=T\int_{BZ}\frac {d^2k}{(2\pi)^2}
\left(1-e^{i{\bf k}\cdot{\bf R}_\mu}\right)
\left[
\frac{1}{\mu k^2}+\frac{1}{\lambda k^2}\right]
\approx \frac{T}{\varepsilon_V}\ln \left(R_\mu/a\right),
\end{equation}
($\varepsilon_V$ is an energy-density scale set by the interaction $V$). 
Note that the fluctuations of a single particle, $\langle u^2\rangle=
\lim_{R_\mu\rightarrow\infty}(\sigma_{\mu0})$ diverges in 2D.
We show here that the logarithmic divergence at large distances is not 
important for the elastic moduli which have dominant contributions from 
fluctuations at small distances.
Consider the 
difference between the renormalized and the zero temperature elastic matrix
\begin{equation}
 \Phi_t^{\alpha\beta}({\bf k})- \Phi_0^{\alpha\beta}({\bf k})= 
n\sum_{\mu\ne 0} (\cos{\bf k}\cdot{\bf R}_\mu-1)
\int\frac {d^2q}{(2\pi)^2} q^\alpha q^\beta V(q) e^{i{\bf q}\cdot{\bf R}_\mu}
\left[ e^{-\frac 1 2 q^2\sigma_{\mu0}}
-1\right].
\end{equation}
Now the Fourier transform only has weight at values of $q<1/R_\mu$, 
so as long as $\sigma_{\mu0}<<R_\mu^2$ we can write
$e^{-\frac 1 2 q^2\sigma_{\mu0}}\approx 1 - \frac 1 2 q^2\sigma_{\mu0}$
and,
\begin{equation}
 \Phi_t^{\alpha\beta}({\bf k})- \Phi_0^{\alpha\beta}({\bf k})\approx 
n\sum_{\mu\ne 0} (\cos{\bf k}\cdot{\bf R}_\mu-1)
\frac 1 2 \sigma_{\mu0} \nabla^2 \partial_\alpha \partial_\beta V(R_\mu)
\end{equation}
Now the $R_\mu$ dependence of $\sigma_{\mu0}$ is logarithmic and 
weak compared to 
the short-range variation in $(\cos{\bf k}\cdot{\bf R}_\mu-1)
\nabla^2 \partial_\alpha \partial_\beta V(R_\mu)$, so short distances will 
dominate and
we are justified to approximate fluctuations by
the nearest-neighbor 
result $\sigma_{\mu0}\approx \langle u^2\rangle_{\rm nn}$. 
Similar arguments apply for the sums that appear in the two-vertex SCHA.
(For logarithmic interactions, $\nabla^2 V(R)=0$, and the 
perturbation theory is not helpful. However, for this case the 
non-perturbative result (\ref{eq:c66ofu2}) derived in Appendix~\ref{ap:log} 
is clearly dominated by the nearest neighbors.)

\section{SCHA calculation of the fluctuation-renormalized elasticity for 
logarithmic interactions in two dimensions}
\label{ap:log}

In this appendix we show how 
to calculate the elastic matrix of a 2D lattice with logarithmic interactions 
within the SCHA as
given by (\ref{eq:effpair}) and demonstrate the non-perturbative dependence
on the size of fluctuations $\langle u^2\rangle$. 
We need the thermal average of the
second derivatives of the interaction potential between each pair of
particles. Assuming the displacement fluctuations to be Gaussian distributed,
this is,
\begin{equation}
 \left\langle 
\frac{\partial^2V({\bf R}_\mu+{\bf u}_\mu-{\bf u}_0)}
{\partial R^\alpha\partial R^\beta}
\right\rangle = \int \frac{d^2r}{Z} 
\frac{\partial^2V({\bf R}_\mu+{\bf r})}
{\partial r^\alpha\partial r^\beta} 
{\cal P}({\bf r})
=\frac{\partial^2}
{\partial R_\mu^\alpha\partial R_\mu^\beta}
 \left\langle V({\bf R}_\mu+{\bf u}_\mu-{\bf u}_0)\right\rangle
\end{equation}
where the derivatives on the RHS do not act on the distribution function,
\begin{equation}
{\cal P}({\bf r})= \exp{\left(-\frac{1}{2}r^\alpha 
(\sigma_{\mu 0}^{-1})^{\alpha\beta}r^\beta\right)}.
\end{equation}
The width of the fluctuations is given by the matrix
\begin{equation}
 \sigma_{\mu 0}^{\alpha\beta}=
\langle[u^\alpha({\bf R}_\mu)-u^\alpha({\bf 0})]
[u^\beta({\bf R}_\mu)-u^\beta({\bf 0})]\rangle, 
\end{equation}
and $Z=2\pi[\hbox{det}(\sigma^{\alpha\beta}_{\mu 0})]^{\frac{1}{2}}$.

We take the isotropic fluctuation approximation,
$\sigma_{\mu 0}^{\alpha\beta}=\delta_{\alpha\beta}\sigma_{\mu 0}$, 
which we find to be
good from numerical evaluation of (\ref{eq:equiofk}) in the HA. 
In this case, because the
potential is circular symmetric, the average will be independent of the angle
$\theta_\mu$ of the ground state position ${\bf R}_\mu$, so
\begin{eqnarray}
 \left\langle V(R_\mu)\right\rangle&=&
\frac{1}{2\pi}\int d\theta_\mu
\int\frac{d^2 r}{Z}V({\bf r}+{\bf R}_\mu)
 e^{-\frac{r^2}{2\sigma_{\mu 0}}}
\\
&=& 
\frac{1}{2\pi}\int d\theta_\mu
\int\frac{d^2 r}{Z}
\int^{R_\mu} dR \frac{d}{dR}
V({\bf r}+{\bf R})
 e^{-\frac{r^2}{2\sigma_{\mu 0}}}\nonumber\\
&=&
\frac{1}{2\pi}\int\frac{d^2 r}{Z}
\int^{R_\mu} \frac{dR}{R} 
\int_{\rho <R}d^2\rho
\nabla_\rho^2
V({\bf r}+{\bf \rho})
 e^{-\frac{r^2}{2\sigma_{\mu 0}}},\nonumber
\end{eqnarray}
where we have made use of the divergence theorem
\hbox{$\int d\theta_\mu(d/dR)=R^{-1}\oint dl\, \hat{\bf n}\cdot{\bf \nabla}=
R^{-1}\int_{\rho<R}d^2\rho\,\nabla_\rho^2$}. We are interested in
the case of
logarithmic interactions, \hbox{$V(R)=-v\ln(R/\xi)$}, 
and we use the
fact that $\nabla^2[\ln(R/\xi)]=2\pi\delta^2({\bf R})$ to give
\begin{equation}
   \left\langle V(R_\mu)\right\rangle
=v \int^{R\mu}\frac{dR}{R}\left[ e^{-R^2/2\sigma_{\mu 0}}-1\right]
= -v\ln(R_\mu/\xi)- v \int^\infty_{R_\mu}\frac{dR}{R}
e^{-R^2/2\sigma_{\mu 0}} + \hbox{const},
\end{equation}
where the integration constant depends on $\sigma_{\mu 0}$ and $\xi$ but not 
on $R_\mu$.
Notice that the correction term to the zero-temperature potential is
non-perturbative in the fluctuation $\sigma_{\mu 0}$. 
The physics of this result is
clear: because the 2D-Laplacian of the interaction potential is zero, Gauss'
theorem applies and the circular symmetric potential gives the same result as
if all of the ``mass'' was at a point in the center of the circle (the ground
state position ${\bf R}_\mu$). 
The non-perturbative correction appears from the
exponentially small tail of the distribution which is outside of the radius
$R_\mu$. This ``shell'' does not contribute to the average interaction, so
it's weight must be subtracted from the bare interaction.
We can easily differentiate the correction term with respect to the lower
limit of the integral, to find
\begin{equation}
\left\langle
 \frac{\partial^2V({\bf R}_\mu+{\bf u}_\mu-{\bf u}_0)}
{\partial R^\alpha\partial R^\beta}
\right\rangle
=\frac{\partial^2V({\bf R}_\mu)} {\partial R_\mu^\alpha\partial
  R_\mu^\beta}
+v\left[\frac{R_\mu^\alpha R_\mu^\beta}{{R_\mu}^4}\left(2+
\frac{{R_\mu}^2}{\sigma_{\mu 0}}\right)
+\frac{\delta_{\alpha\beta}}{{R_\mu}^2}\right]
e^{-\frac{{R_\mu}^2}{2\sigma_{\mu 0}}}.
\end{equation}
This expression should be inserted into (\ref{eq:effpair}) to give the
renormalized elastic matrix. It is then straightforward to derive the
softening of the shear modulus within the SCHA as given in
Eq.~(\ref{eq:c66ofu2}) and Fig.~\ref{fig:2D_c66ofu2}.

\section{Equivalence of elastic response
with inverse propagator}
\label{ap:greens}

We demonstrate in this appendix that the elastic response to an external force 
is given by the inverse propagator (or Green's function). In the presence of a 
force ${\bf f}_\mu$ acting on each particle, the Hamiltonian is changed to,
\begin{equation}
  H_f[{\bf u}_\mu]=H[{\bf u}_\mu] -\sum_\mu {\bf f}_\mu\cdot {\bf u}_\mu.
\end{equation}
The definition of elastic response within the thermally fluctuating region
is that the force is given by Hooke's law,
\begin{equation}
  f_\mu^\alpha=\sum_\nu 
\Phi^{\alpha\beta}({\bf R}_\nu-{\bf R}_\mu) U_\nu^\beta,
\end{equation}
which is valid to lowest order in the mean displacements
$U_\nu^\beta=\langle u_\nu^\beta\rangle_f$. The displacements are calculated
with straightforward statistical mechanics,
\begin{eqnarray}
  U_\nu^\beta&=& Z_f^{-1}\int\prod_\mu d^2u_\mu\,  u_\nu^\beta\,
e^{-\left(H[{\bf u}_\lambda]-\sum_{\lambda}{\bf f}_{\lambda}\cdot 
{\bf u}_{\lambda}\right)
/T},\\
&=&
\frac{1}{T}\sum_\mu f_\mu^\alpha \langle u_\nu^\beta u_\mu^\alpha\rangle_{f=0}
+{\cal O}(f^2).
\end{eqnarray}
If we have lattice-translational symmetry, the Fourier transform of this reads 
as $U^\beta_{\bf k}= f_{\bf k}^\alpha G^{\alpha\beta}({\bf k})$, where the
Green's function is defined by 
$\langle u_{\bf k}^\beta u_{{\bf k}'}^\alpha\rangle_{f=0}
=TG^{\alpha\beta}({\bf k})\delta({\bf k}+{\bf k}')$, so that the elastic
matrix in ${\bf k}$-space is exactly the inverse Green's function,
\begin{equation}
  \Phi^{\alpha\beta}({\bf k})=(G^{-1})^{\alpha\beta}({\bf k}).
\end{equation}
Note that this example of the classical
fluctuation-dissipation theorem goes beyond the
equipartition result, which is only valid for the harmonic approximation. The
true elastic response of the fluctuating system is always given by the inverse 
of the propagator.

If we construct the free energy corresponding to $H_f$, we will find the
result,
\begin{equation}
  F_f=F_{f=0}-\frac{1}{2}\int_{\rm BZ} \frac{d^dk}{(2\pi)^d}
(G^{-1})^{\alpha\beta}({\bf k}) U^\alpha_{\bf k} U^\beta_{-{\bf k}}
+{\cal O}(f^3).
\end{equation}
This is for a system with forces specified. However, one is often interested
in the case where a certain displacement is enforced (e.g. when boundary
conditions are changed in a numerical simulation). We must then make a
Legendre transformation to the relevant thermodynamic potential that 
depends on a fixed displacement,
\begin{eqnarray}
  A(T,[{\bf U}_\mu])&=&F_f +\sum_\mu f_\mu^\alpha  U_\mu^\alpha\\
&=& 
F_{f=0}+\frac{1}{2}\int_{\rm BZ} \frac{d^dk}{(2\pi)^d}
(G^{-1})^{\alpha\beta}({\bf k}) U^\alpha_{\bf k} U^\beta_{-{\bf k}}.
+{\cal O}(U^3).
\end{eqnarray}
With this formulation, we see that the elastic matrix may be found as the
second derivative of the thermodynamic potential with fixed displacements,
$\Phi^{\alpha\beta}({\bf k})=\left.\partial^2A/\partial U_{\bf k}^\alpha
\partial U_{-{\bf k}}^\beta\right|_{U=0}$. For example, the $k\rightarrow 0$
shear modulus may be measured as the response to the change in the angle
$\theta$ determining the boundaries of a finite system as in
Ref.~\onlinecite{Cai}. 
As the Hamiltonian will depend on the ``displacement'' $\theta$, we can
differentiate the relevant free energy, $A(T,\theta)$, to give,
\begin{equation}
  \mu=L^{-d}\left[
\left\langle \frac{\partial^2H}{\partial\theta^2}\right\rangle_{\theta=0}
-\frac{1}{T}  
\left\langle \left(\frac{\partial H}{\partial\theta}\right)^2\right
\rangle_{\theta=0}
+\frac{1}{T}  
\left\langle \frac{\partial H}{\partial\theta}\right\rangle_{\theta=0}^2
\right]
\end{equation}
The important point is that this shear modulus is exactly the same as the
shear modulus found from the $k\rightarrow 0$ limit of the inverse of the
transverse propagator $\mu=\lim_{k_y\rightarrow 0}[(G^{-1})^{xx}(k_y)/k_y^2]$.

\section{The (3+4)-SCA for pairwise interactions}
\label{ap:34}

In this appendix we show how to explicitly calculate the (3+4) diagrammatic
expansion of Fig.~\ref{fig:diags_34}, for the case of pairwise interactions.
For a Hamiltonian of the form (\ref{eq:pairwise}),
the third and fourth rank anharmonic tensors defined by Eqs.~(\ref{eq:Hm}) 
and~(\ref{eq:AtoPhi}) will be given by,
\begin{eqnarray}
  \Phi^{\lambda_1\lambda_2\lambda_3}({\bf k}_1,{\bf k_2})
&=&n\sum_{\mu\ne0}
\left(e^{i{\bf k}_1.{\bf R}_\mu}+
e^{i{\bf k}_2.{\bf R}_\mu}+
e^{-i({\bf k}_1+{\bf k}_2).{\bf R}_\mu}\right)
\frac{\partial^3V}{\partial R_\mu^{\lambda_1}\partial R_\mu^{\lambda_2}
\partial R_\mu^{\lambda_3}}\nonumber\\
&=&in^2\sum_{{\bf Q}_\mu} f_3({\bf Q}_\mu+{\bf k}_1)+
 f_3({\bf Q}_\mu+{\bf k}_2)+
 f_3({\bf Q}_\mu-{\bf k}_1-{\bf k}_2),
\end{eqnarray}
and
\begin{eqnarray}
  \Phi^{\lambda_1\lambda_2\lambda_3\lambda_4}({\bf k}_1,{\bf k_2},{\bf k}_3)
&=&n\sum_{\mu\ne0}
\left(1-e^{i{\bf k}_1.{\bf R}_\mu}\right)
\left(1-e^{i{\bf k}_2.{\bf R}_\mu}\right)
\left(1-e^{i{\bf k}_3.{\bf R}_\mu}\right)
\frac{\partial^4 V}{\partial R_\mu^{\lambda_1}\partial R_\mu^{\lambda_2}
\partial R_\mu^{\lambda_3}\partial R_\mu^{\lambda_4}}\nonumber\\
&=&n^2\sum_{{\bf Q}_\mu} f_4({\bf Q}_\mu)
 -f_4({\bf Q}_\mu-{\bf k}_1)
 -f_4({\bf Q}_\mu-{\bf k}_2)
 -f_4({\bf Q}_\mu-{\bf k}_3)
 -f_4({\bf Q}_\mu-{\bf k}_1-{\bf k}_2-{\bf k}_3)\nonumber\\&&\hspace{1cm}
 +f_4({\bf Q}_\mu-{\bf k}_1-{\bf k}_2)
 +f_4({\bf Q}_\mu-{\bf k}_2-{\bf k}_3)
 +f_4({\bf Q}_\mu-{\bf k}_3-{\bf k}_1)
,
\end{eqnarray}
with $f_m({\bf Q})=Q_{\lambda_1}\ldots Q_{\lambda_m}V({\bf Q})$.
The above 
Poisson resummations to reciprocal-lattice sums allow for more convenient
evaluation, especially in the case of long-ranged interactions.
The contribution to the self energy from the ``flying-saucer'' diagram is,
\begin{equation}\label{eq:flyingsaucer}
  \Sigma^{\lambda_1\lambda_2}_3({\bf k},G)=
\frac{L^d}{2T}
\left\langle 
\frac{\delta^2 ({H_3}^2)}{\delta u^{\lambda_1}_{\bf k}
\delta u^{\lambda_2}_{-{\bf k}}}
\right\rangle_{G}
=-\frac{T}{2}\int_{\rm BZ}\frac{d^dk_1}{(2\pi)^d}
G^{\lambda_3\lambda_4}({\bf k}_1)
G^{\lambda_5\lambda_6}({\bf k}+{\bf k}_1)
\Phi^{\lambda_1\lambda_3\lambda_5}({\bf k},{\bf k}_1)
\Phi^{\lambda_2\lambda_4\lambda_6}({\bf k},{\bf k}_1),
\end{equation}
while the contribution from the first, ``Hartree''-type diagram is
\begin{equation}\label{eq:hartree}
  \Sigma^{\lambda_1\lambda_2}_4({\bf k},G)
=-L^d
\left\langle 
\frac{\delta^2 H_4}{\delta u^{\lambda_1}_{\bf k}
\delta u^{\lambda_2}_{-{\bf k}}}
\right\rangle_{G}
=-\frac{T}{2} \int_{\rm BZ} \frac{d^dk_1}{(2\pi)^d}
G^{\lambda_3\lambda_4}({\bf k_1})\Phi^{\lambda_1\lambda_2\lambda_3\lambda_4}
({\bf k},-{\bf k},{\bf k}_1).
\end{equation}
%

\section{The Two-Vertex-SCHA for pairwise interactions}
\label{ap:n=2}

We now describe the details of the two-vertex-SCHA of Section~\ref{sec:new}
for a system with pairwise interactions. Splitting the effective elastic
matrix of Eq.~(\ref{eq:neweff}) term by term we have,
\begin{equation}\label{eq:split}
  \Phi_t^{\alpha\beta}({\bf k})=
  \Phi_1^{\alpha\beta}({\bf k})+
  \Phi_2^{\alpha\beta}({\bf k})+
  \Phi_{2c}^{\alpha\beta}({\bf k}),
\end{equation}
with $\Phi_1^{\alpha\beta}({\bf k})$ given by the SCHA
formula~(\ref{eq:effpair}), 
\begin{equation}\label{eq:phi2real}
  \Phi_2^{\alpha\beta}({\bf k})=-\frac{L^d}{T}
\left\langle
 \frac{\delta H}{\delta u^\alpha_{-\bf k}}
 \frac{\delta H}{\delta u^\beta_{\bf k}}
\right\rangle
= \frac{-1}{4TL^d}\sum_{\scriptsize\begin{array}{l} 
   \mu\ne\nu\\
   \rho\ne\sigma
    \end{array}}
\left( e^{i{\bf k}\cdot{\bf R}_\mu}- e^{i{\bf k}\cdot{\bf R}_\nu}\right)
\left( e^{-i{\bf k}\cdot{\bf R}_\rho}- e^{-i{\bf k}\cdot{\bf R}_\sigma}\right)
\left\langle
\left.\frac{\partial V}{\partial r^\alpha}\right|_{\mu-\nu}
\left.\frac{\partial V}{\partial r^\beta}\right|_{\rho-\sigma}
\right\rangle,
\end{equation}
and,
\begin{equation}
   \Phi_{2c}^{\alpha\beta}({\bf k})=
\frac{1}{TL^d}\Phi_1^{\alpha{\lambda_1}}({\bf k})
\Phi_1^{\beta{\lambda_2}}({\bf k})
\left\langle
u_{-{\bf k}}^{\lambda_1}
u_{{\bf k}}^{\lambda_2}
\right\rangle,
\end{equation}
where the averages are taken with respect to the effective elastic
Hamiltonian. 
In the independent fluctuation approximation, we can write the entire elastic
matrix as a function of $\langle u^2\rangle$ and $T$ (note that in 2D, although
$\langle u^2\rangle$ diverges, we can take
the nearest-neighbor value $\langle u^2\rangle_{\rm nn}$ as the 
sum in (\ref{eq:phi2real}) is dominated by short distances),
\begin{equation}\label{eq:phi2}
   \Phi_2^{\alpha\beta}({\bf k})=\frac{n^3}{T}\!
\sum_{{\bf Q},{\bf Q}'}
\left[ g^{\alpha\beta}_{{\bf Q} ,  {\bf Q}'}
-g^{\alpha\beta}_{{\bf Q}-{\bf k} ,  {\bf Q}'}
-g^{\alpha\beta}_{{\bf Q} ,  {\bf Q}'+{\bf k}}
+g^{\alpha\beta}_{{\bf Q}-{\bf k} ,  {\bf Q}'+{\bf k}}
\right]
+\frac{n^2}{2T}\!
\int \!\!\! \frac{d^dq'}{(2\pi)^d}\! \sum_{\bf Q}
\left[
2 h^{\alpha\beta}_{{\bf Q}-{\bf q}',{\bf q}'}
- h^{\alpha\beta}_{{\bf Q}-{\bf q}'-{\bf k},{\bf q}'}
- h^{\alpha\beta}_{{\bf Q}-{\bf q}'+{\bf k},{\bf q}'}
\right],
\end{equation}
with
\begin{equation}
  g^{\alpha\beta}_{{\bf q}_1, {\bf q}_2}=
q_1^\alpha q_2^\beta V(q_1)V(q_2)
e^{-(q_1^2+q_2^2)\langle u^2\rangle/d}
\left[
e^{-{\bf q}_1.{\bf q}_2\langle u^2\rangle/d}
-1
\right],
\end{equation}
and
\begin{equation}
  h^{\alpha\beta}_{{\bf q}_1, {\bf q}_2}=
q_1^\alpha q_2^\beta V(q_1)V(q_2)
e^{-(q_1^2+q_2^2)\langle u^2\rangle/d}
\left[
e^{-{\bf q}_1.{\bf q}_2\langle u^2\rangle/d}
-1
\right]^2.
\end{equation}
Finally the correction term is given by,
\begin{equation}
   \Phi_{2c}^{\alpha\beta}({\bf k})=
\frac{\langle u^2\rangle }{dnT}\,
\Phi_1^{\alpha\lambda}({\bf k})\Phi_1^{\beta\lambda}({\bf k}).
\end{equation}
To complete the  two-vertex-SCHA one must find a self-consistent solution 
numerically for $\langle u^2\rangle$ by inserting the effective elastic 
matrix in the equipartition result~(\ref{eq:equi}).

\section{Modifications for interacting elastic strings in three dimensions}
\label{ap:string}

We now outline the generalization of the self-consistent methods to the
problem of elastic strings directed along $z$ and interacting  within 2D
planes of equal $z$. First, the harmonic energy is given by,
\begin{equation}
  H_2=\frac{1}{2}\int_{\rm BZ}\frac{d^2k_\perp}{(2\pi)^2}
\int_{-\infty}^{\infty} \frac{dk_z}{2\pi}
\left[
\Phi_{0,\,2\rm D}^{\alpha\beta}({\bf k}_\perp)+
\delta_{\alpha\beta}n\epsilon_l k_z^2
\right]
u_{\bf k}^\alpha u_{-{\bf k}}^\beta.
\end{equation}
The generalization of Eq.~(\ref{eq:flyingsaucer}) is,
\begin{equation}\label{eq:flyingsaucerstr}
  \Sigma^{\lambda_1\lambda_2}_3({\bf k},G)
=-\frac{T}{2}\int_{\rm BZ}\frac{d^2k_{\perp}'}{(2\pi)^2}
\int \frac{dk_{z}'}{2\pi}
G^{\lambda_5\lambda_6}({\bf k}+{\bf k}')
G^{\lambda_3\lambda_4}({\bf k}')
\Phi_{2\rm D}^{\lambda_1\lambda_3\lambda_5}({\bf k}_\perp,{\bf k}_{\perp}')
\Phi_{2\rm D}^{\lambda_2\lambda_4\lambda_6}({\bf k}_\perp,{\bf k}_{\perp}'),
\end{equation}
and of Eq.~(\ref{eq:hartree}) is,
\begin{equation}\label{eq:hartreestr}
  \Sigma^{\lambda_1\lambda_2}_4({\bf k},G)
=-\frac{T}{2} \int_{\rm BZ} \frac{d^2k_\perp'}{(2\pi)^2}
 \int \frac{dk_z'}{2\pi}
G^{\lambda_3\lambda_4}({\bf k}')
\Phi_{2\rm D}^{\lambda_1\lambda_2\lambda_3\lambda_4}
({\bf k}_\perp,-{\bf k}_\perp,{\bf k}_\perp').
\end{equation}
The matrices with $2D$ subscripts are to be calculated for the interactions
within a given 2D plane.

For the two-vertex-SCHA we now have to be careful to include the
$z$-correlations in the fluctuation terms, $\langle u^2(z)\rangle=\langle
u^\alpha(z)u^\alpha(0)\rangle$. We
can again split the effective elastic matrix as in
(\ref{eq:split}), with $\Phi_1$ being the 2D SCHA elastic modulus plus the
single string elasticity, and
\begin{eqnarray}\label{eq:3dphi2}
   \Phi_{2}^{\alpha\beta}({\bf k})&=&\frac{n^3}{T}
\sum_{{\bf Q},{\bf Q}'}
\int^\infty_{-\infty}dz \,\,e^{-ik_zz}
\left[ g^{\alpha\beta}_{{\bf Q} ,  {\bf Q}'}(z)
-g^{\alpha\beta}_{{\bf Q}-{\bf k} ,  {\bf Q}'}(z)
-g^{\alpha\beta}_{{\bf Q} ,  {\bf Q}'+{\bf k}}(z)
+g^{\alpha\beta}_{{\bf Q}-{\bf k} ,  {\bf Q}'+{\bf k}}(z)
\right]\\
&&+\frac{n^2}{2T}\int \frac{d^2q'}{(2\pi)^2}\sum_{\bf Q}
\int^\infty_{-\infty}dz  \,\,e^{-ik_zz}
\left[
2 h^{\alpha\beta}_{{\bf Q}-{\bf q}',{\bf q}'}(z)
- h^{\alpha\beta}_{{\bf Q}-{\bf q}'-{\bf k},{\bf q}'}(z)
- h^{\alpha\beta}_{{\bf Q}-{\bf q}'+{\bf k},{\bf q}'}(z)
\right],\nonumber
\end{eqnarray}
with
\begin{equation}
  g^{\alpha\beta}_{{\bf q}_1, {\bf q}_2}(z)=
q_1^\alpha q_2^\beta V(q_1)V(q_2)
e^{-\frac 12 (q_1^2+q_2^2)\langle u^2(0)\rangle}
\left[
e^{-\frac 12 {\bf q}_1.{\bf q}_2\langle u^2(z)\rangle}
-1
\right],
\end{equation}
and
\begin{equation}
  h^{\alpha\beta}_{{\bf q}_1, {\bf q}_2}(z)=
q_1^\alpha q_2^\beta V(q_1)V(q_2)
e^{-\frac 12 (q_1^2+q_2^2)\langle u^2(0)\rangle}
\left[
e^{-\frac 12 {\bf q}_1.{\bf q}_2\langle u^2(z)\rangle}
-1
\right]^2,
\end{equation}
where $\langle u^2(z)\rangle$ is defined in Eq.~(\ref{eq:u2ofz}).
The last, correction term is given by,
\begin{equation}
  \Phi_{2c}^{\alpha\beta}({\bf k})=
\frac{1}{2nT}\Phi_{1,\,{\rm 2D}}^{\alpha\lambda}({\bf k}_\perp)
\Phi_{1.\,2D}^{\lambda\beta}({\bf k}_\perp)
\int dz\, e^{-ik_zz}
\langle u^2(z)\rangle.
\end{equation}

\begin{multicols}{2}
\narrowtext

\end{multicols}


\begin{thebibliography}{99}
\bibitem{GSE} We ignore the
cohesion energy of the ground state as it does not influence the fluctuations
considered in this paper.
However, if we wanted to calculate the full free energy, e.g. to compare to
the liquid state, we would need to include this ground state energy.
\end{thebibliography}
\end{document}